\documentclass[fleqn,usenatbib]{mnras}

\usepackage{newtxtext,newtxmath}

\usepackage[normalem]{ulem}

\usepackage{adjustbox}
\usepackage[T1]{fontenc}

\DeclareRobustCommand{\VAN}[3]{#2}
\let\VANthebibliography\thebibliography
\def\thebibliography{\DeclareRobustCommand{\VAN}[3]{##3}\VANthebibliography}

\usepackage{graphicx}	% Including figure files
\usepackage{amsmath}	% Advanced maths commands
\usepackage{siunitx}
\usepackage{multicol}
\usepackage{footnote}
\usepackage{threeparttable}

\usepackage{xspace}

\newcommand{\hestia}{\textsc{hestia}\xspace}

\title[A gas view of the MW-M31 system]{Cold and Hot gas distribution around the Milky-Way -- M31 system in the HESTIA simulations}

\author[M. Damle]{
Mitali Damle$^{1}$\thanks{E-mail: damle@uni-potsdam.de}, Martin Sparre$^{1,2}$,
Philipp Richter$^{1}$,
Maan H. Hani$^{3}$\thanks{***Herschel fellow***},
Sebasti\'an~E.~Nuza$^{4,5}$,
\newauthor
\hspace{0.01cm} Christoph Pfrommer$^{2}$,
Robert J. J. Grand$^{6,11,12}$,
Yehuda Hoffman$^{9}$,
Noam Libeskind$^{2}$,
Jenny G. Sorce$^{2, 10}$,
\newauthor
\hspace{0.01cm} Matthias Steinmetz$^{2}$,
Elmo Tempel$^{7,13}$,
Mark Vogelsberger$^{8}$,
Peng Wang$^{2}$
\\
\\
$^{1}$Institut für Physik und Astronomie, Universität Potsdam, Campus Golm, Haus 28, Karl-Liebknecht Straße 24-25, 14476 Potsdam, Germany\\
$^{2}$Leibniz-Institut für Astrophysik Potsdam (AIP), An der Sternwarte 16, 14482 Potsdam, Germany\\
$^{3}$Department of Physics and Astronomy, McMaster University, Hamilton Ontario L8S 4M1, Canada\\
$^{4}$Instituto de Astronom\'{\i}a y F\'{\i}sica del Espacio (IAFE, CONICET-UBA), CC 67, Suc. 28, 1428 Buenos Aires, Argentina\\
$^{5}$Facultad de Ciencias Exactas y Naturales (FCEyN), Universidad de Buenos Aires (UBA), Buenos Aires, Argentina\\
$^{6}$Max-Planck-Institut für Astrophysik, Karl-Schwarzschild-Str 1, D-85748 Garching, Germany\\
$^{7}$Tartu Observatory, University of Tartu, Observatooriumi 1, 61602 Tõravere, Estonia\\
$^{8}$Department of Physics, Kavli Institute for Astrophysics and Space Research, Massachusetts Institute of Technology, Cambridge, MA 02139, USA\\
$^{9}$Racah Institute of Physics, Hebrew University, Jerusalem, 91904, Israel\\
$^{10}$Univ Lyon, ENS de Lyon, Univ Lyon1, CNRS, Centre de Recherche Astrophysique de Lyon UMR5574, F-69007, Lyon, France\\
$^{11}$Instituto de Astrof\'isica de Canarias, Calle Vía L\'actea s/n, E-38205 La Laguna, Tenerife, Spain\\
$^{12}$Departamento de Astrof\'isica, Universidad de La Laguna, Av. del Astrof\'isico Francisco S\'anchez s/n, E-38206, La Laguna, Tenerife, Spain\\
$^{13}$Estonian Academy of Sciences, Kohtu 6, 10130 Tallinn, Estonia\\
}

\date{Accepted XXX. Received YYY; in original form ZZZ}

\pubyear{2021}

\begin{document}
\label{firstpage}
\pagerange{\pageref{firstpage}--\pageref{lastpage}}
\maketitle

\begin{abstract}

Recent observations have revealed remarkable insights into the gas reservoir in the circumgalactic medium (CGM) of galaxy haloes. In this paper, we characterise the gas in the vicinity of Milky Way and Andromeda analogues in the \hestia (High resolution Environmental Simulations of The Immediate Area) suite of constrained Local Group (LG) simulations. The \hestia suite comprise of a set of three high-resolution {\sc arepo}-based simulations of the LG, run using the Auriga galaxy formation model. For this paper, we focus only on the $z = 0$ simulation datasets and generate mock skymaps along with a power spectrum analysis to show that the distributions of ions tracing low-temperature gas (H{\,\sc i} and Si{\,\,\sc iii}) are more clumpy in comparison to warmer gas tracers (O{\,\sc vi}, O{\,\sc vii} and O{\,\sc viii}). We compare to the spectroscopic CGM observations of M31 and low-redshift galaxies. \hestia under-produces the column densities of the M31 observations, but the simulations are consistent with the observations of low-redshift galaxies. A possible explanation for these findings is that the spectroscopic observations of M31 are contaminated by gas residing in the CGM of the Milky Way.
\end{abstract}


%%%%%%%%%%%%%%%%%%%%%%%%%%%%%%%%%%%%%%%%%%%%%%%%%%

%%%%%%%%%%%%%%%%% BODY OF PAPER %%%%%%%%%%%%%%%%%%

\section{Introduction}
\label{section: Introduction}

Our understanding of the tenuous gas reservoir surrounding galaxies, better known as the circumgalactic medium (CGM), has dramatically improved since its first detection, back in the 1950s \citep{spitzer1956possible, munch1961interstellar, bahcall1969absorption}. The CGM is a site through which pristine, cold intergalactic medium (IGM) gas passes on its way into the galaxy and it is also the site where metal-enriched gas from the interstellar medium (ISM) gets dumped via outflows and winds \citep{2017MNRAS.470.4698A,2019MNRAS.483.4040S}. CGM gas is often extremely challenging to detect in emission due to its low column densities. Therefore, most of our knowledge about its nature stems from absorption line studies \citep{werk2014cos, tumlinson2017} of quasar sightlines passing through the CGM of foreground galaxies.

Observational datasets from instruments like Far Ultraviolet Spectroscopic Explorer (FUSE, see \citealt{Moos_2000,Savage_2000,Sembach_2000}), Space Telescope Imaging Spectrograph (HST-STIS, see \citealt{Woodgate_1998, Kimble1998}) and Cosmic Origins Spectrograph (HST-COS, see \citealt{froning2009cosmic, green2011cosmic}) have revolutionised our understanding of not just the MW CGM but the CGMs of other galaxies as well \citep{richter2001diversity, lehner2012MNRAS.424.2896L, herenz2013A&A...550A..87H, Tumlinson_2013, werk2013, fox2014cos, werk2014cos, Richter_2017}.

Numerous studies through the last decade involving quasar absorption line studies of various low and intermediate ions tracing a substantial range in temperatures and densities have revealed the complex, multiphase structure of the CGM \citep{Nielsen_2013, Tumlinson_2013, bordoloi2014cos, Richter_2016, Lehner_2018}.
\citet{lehner2020project} have gone a step ahead in quasar absorption line studies by obtaining multi-ion deep observations of several sightlines heterogeneously piercing the CGM of a single galaxy (M31).

Recent studies conclude that a significant percentage of galactic baryons could lie in the warm-hot virialized gas phase \citep{peeples2014budget, tumlinson2017}, increasingly emphasizing the importance of high ions in describing the CGM mass budget \citep{tumlinson2017}. O{\,\sc vi}, which is an important tracer of the warm-hot CGM ($T \sim 10^{5.5}$ K), has been detected in gas reservoirs around star forming galaxies in Far UV \citep{tumlinson2011large}. Even hotter CGM gas, traced primarily by O{\,\sc vii} and O{\,\sc viii}, has been detected around galaxies in X-ray studies \citep{das2019evidence, das2020detection}. Apart from these high ions, Coronal Broad Lyman alpha absorbers could also contribute towards constituting the hot CGM \citep{richter2020ApJ...892...33R}.

Significant progress is also being made via systematic CGM studies targeting diverse galaxy samples which provide insightful views into the synergy between the CGM and the evolution of its host galaxy. The presence of warm gas clouds around late-type galaxies at low redshift \citep{stocke2013characterizing}, the impact of starbursts \citep{borthakur2013impact} and AGN \citep{berg2018}, evidence of a bimodal metallicity distribution in the form of metal-poor, pristine and metal-rich, recycled gas streams \citep{lehner2013bimodal} have given us a peek into the interplay between the CGM and its parent galaxy. The theory of galactic winds injecting metal-rich gas from the ISM out to the CGM \citep{hummels2013constraints, ford2013MNRAS.432...89F} is now being supported by observational evidence \citep{rupke2019100}.

Despite all the advancements in the past few years, limited sightline observations and our technological inability to probe substantially lower column densities in the CGM of other galaxies indicate that we cannot yet fully rely solely on these studies to give us a complete picture of the workings of the CGM \citep{tumlinson2017}. Therefore, studying the MW and the LG CGM (which will always have better CGM datasets as compared to those for non-LG galaxies) assumes a great importance in this context.

High-velocity, warm O{\,\sc vi} gas has been observed extensively around the MW \citep{sembach2003highly, savage2003distribution, wakker2003far}. HST UV spectra of a list of low and intermediate ions have further helped us track the expanse of high-velocity clouds (HVCs) around our galaxy \citep{lehner2012MNRAS.424.2896L, herenz2013A&A...550A..87H}. A low-velocity, cool-ionized CGM component has also been detected recently around our MW \citep{zheng2019revealing, bish2020quastar}. Additionally, a long hypothesised hot, diffuse galactic gas phase \citep{gupta2012huge} has been observed using the highly ionized O{\,\sc vii} and O{\,\sc viii} ions \citep{miller2015constraining, das2019discovery}. While the observations of our own galaxy's CGM certainly provide us with more sightlines and enable us to detect slightly lower column densities as compared to other galaxies' CGMs, galactic CGM observations are fraught with a greater possibility of contamination from sources lying in the line-of-sight of our observations, thereby masking the true nature of our galaxy's CGM.

With the advent of the above observations, complementary studies with regards to the CGMs around the galaxies, generated using cosmological galaxy formation simulations, started gaining momentum \citep{vogelsberger2020NatRP...2...42V}. Cosmological simulations, in general, have been extremely successful in replicating many pivotal observational properties central to the current galaxy formation and evolution model \citep{vogels2014MNRAS.444.1518V,vogelsberger2014Natur.509..177V}. These include galaxy morphologies \citep{ceverino2010high, aumer2013towards, marinacci2014formation, somerville2015physical, grand2017auriga}, galaxy scaling relations \citep{booth2009cosmological, Angulo_2012, vogelsberger2013model}, $M_*/M_\text{halo}$ relationship \citep{Behroozi_2010, 2013MNRAS.428.3121M}, and star formation in galaxies \citep{behroozi2013average, Agertz_2015, 2015MNRAS.447.3548S,2015MNRAS.450.4486F,2017MNRAS.466...88S,2019MNRAS.485.4817D}. Like observations, cosmological simulations provide different approaches to quantify the typical baryon and metal budgets of galaxies \citep{ford2014tracing, schaye2015eagle, Suresh_2016, Hani_2019, tuominen2020eagle}. They reveal how the motions of gas manifests itself in various forms like inflow streams from the IGM, or replenished outflows from the galaxy out to its CGM, or stellar winds or supernovae and AGN feedback \citep{2019MNRAS.490.3234N,2021arXiv210210913W,2021arXiv210210126A}.

Given that the computational studies of the CGM have provided an enormous insight into the evolution of galaxies, it is worthwhile to look back to our local environment, i.e. the Local Group (LG). Apart from tracking the formation history of MW-M31 \citep{ibata2013vast, hammer2013vast, scannapieco2015A&A...577A...3S} and the accretion histories of MW-like galaxies \citep{nuza2019}, our LG, over the past decade, has proved to be an ideal site for studies involving $\Lambda$CDM model tests \citep{klypin1999missing, wetzel2016reconciling, lovell2017properties}, dwarf galaxy formation and evolution \citep{tolstoy2009star, garrison2014too, pawlowski2017lopsidedness, samuel2020profile}, effects of environment on star formation histories of MW-like galaxies \citep{creasey2015ApJ...800L...4C}, local universe re-ionization \citep{ocvirk2020}, and the cosmic web \citep{nuza2014MNRAS.445..988N, forero2015local, metuki2015galaxy}. Observational constraints of the Local Universe have resulted in an emergence of constrained simulations, where the large-scale structure resembles the observations \citep{nuza2010, libeskind2011, knebe2011, dicintio2013, nuza2013}. It is also worthwhile to note that such LG constrained simulations might be the setups best equipped to separate out any sources of possible contamination towards the MW CGM.

A simulation of a Milky-Way-like galaxy in a constrained environment was done by the CLUES (Constrained Local UniversE Simulations) project \citep{gottlober2010constrained}, which were one of the first cosmological simulations to include a realistic local environment within the large-scale LG structure. \citet{nuza2014distribution} carried out a study on the $z = 0$ gas distribution around MW and M31 in the CLUES simulation to characterize the effect of cosmography on the LG CGM. They analysed the cold and hot gas phases, computed their masses and accretion/ejection rates, and later compared their results with the absorption-line observations from \cite{Richter_2017}.

We build upon the approach adopted in \citet{nuza2014distribution} by analysing the constrained LG simulations, {\hestia} \citep{libeskind2020hestia}, which in comparison to the original CLUES simulations have better constrained initial conditions. In \hestia we, furthermore, use the Auriga galaxy formation model \citep{grand2017auriga}, which produces realistic Milky-Way-mass disc galaxies. In comparison to the previous CLUES simulations, we carry out a more extensive analysis to predict column densities of a range of tracer ions (H{\,\sc i}, Si{\,\sc iii}, O{\,\sc vi}, O{\,\sc vii} and O{\,\sc viii}) selected to give a complete view of the various gas phases in and around the galaxies. This helps us, for example, with the interpretation of absorption studies of the LG CGM gas.
 
The aim of this paper is to provide predictions for absorption-line observations of the gas in the LG. We achieve this by studying the gas around LG galaxies in the state-of-the-art constrained magnetohydrodynamical (MHD) simulations, \hestia (High resolution Environmental Simulations of The Immediate Area). The comparison between \hestia and some of the recent observations makes it possible to constrain the galaxy formation models of our simulations. 

This paper is structured as follows: §~\ref{section: Methods} describes the analysis tools and the simulation. We present our results in §~\ref{section: Results}, which include Mollweide projection maps (§\ref{Skymaps}), power spectra (§\ref{PowerSpec}) and radial column density profiles (§\ref{ColDens}). We compare our results with some of the recent observations and other simulations in §\ref{Comparisons} and \ref{SimComparisons}. Further, we discuss the implications of our results in the context of current theories about CGM and galaxy formation and evolution in §\ref{Discussion}. We also analyse the possibility of MW's CGM gas interfering with M31's CGM observations in §\ref{MWcontamination}. Finally, we sum up our conclusions and provide a quick note about certain caveats and ideas to be implemented in future projects (§\ref{Conclusions}).

\section{Methods}
\label{section: Methods}

We use three high resolution realizations from the \hestia suite, a set of intermediate and high resolution cosmological magneto-hydrodynamical constrained simulations of the LG, analysed only at the present time ($z=0$). The \hestia project is a part of the larger CLUES collaboration \citep{gottlober2010constrained, Libeskind2010, Sorce2016a, Carlesi2016}, whose principal aim is to generate constrained simulations of the local universe in order to match the mock observational outputs with real observations from our galactic neighbourhood. 

The following subsection summarises the technical specifications of these simulations. A more extensive description of the simulations can be found in the official \hestia release paper \citep{libeskind2020hestia}. 

\subsection{Simulations}
\label{subsection: Simulations}

\subsubsection{Initial Conditions}
\label{subsubsection: Initial Conditions}

The small scale initial conditions are obtained from a sampling of the peculiar velocity field. The CosmicFlows-2 catalog \citep{tully2013cosmicflows}, used to derive peculiar velocities, provides constraints up to distances farther than that was available for the predecessor CLUES simulation. Reverse Zel'dovich technique \citep{doumler2013} handles the cosmic displacement field better, hence offering smaller structure shifts. A new technique, bias minimisation scheme \citep{sorce2016}, has been employed for \hestia simulations to ensure that the LG characteristic objects (e.g. Virgo cluster) have proper mass. The above mentioned new elements (see \citealt{Sorce2016a} for further details) in conjunction with the earlier aspects of constrained realization \citep{hoffman1991constrained} and Wiener Filter \citep{sorce2014} offer \hestia a clear edge over the previous generation CLUES simulations. 

Low-resolution, constrained, dark-matter only simulations are the fields from which halo pairs resembling our LG were picked up for intermediate and high resolution runs. Note that only the highest resolution realizations (those labelled 09$-$18, 17$-$11 and 37$-$11) are used for our analysis in this paper. The first and second numbers in the simulation nomenclature represent the seed for long and short waves, respectively, both of which together constitute to the construction of the initial conditions. Two overlapping $2.5h^{-1}$ Mpc spheres centred on the two largest $z = 0$ LG members (MW and M31) represent the effective high resolution fields which are populated with ${8192^3}$ effective particles. The mass resolution for the DM particles (gas cells) in the high-resolution simulations is $1.5\times {10^5}$ M$_{\odot}$ ($2.2\times {10^4}$ M$_{\odot}$), while the softening length ($\epsilon$) for the DM is 220 pc.

While the entire process of selecting cosmographically correct halo pairs involves handpicking MW-M31 candidates with certain criteria (halo mass, separation, isolation) that lie within the corresponding observational constraints, there are yet a few other bulk parameters ($M_*$ vs $M_\text{halo}$, circular velocity profile) and dynamical properties (total relative velocities) which are organically found to agree well with observations \citep{guo2010galaxies,van2012m31,mcleod2017estimating}.

\subsubsection{Galaxy formation model}
\label{subsubsection:galformmodel}\label{subsubsection: MHD}

\begin{table*}
\caption{Properties of MW and M31 analogues at $z=0$ for the three LG \hestia simulations. The simulations are referred to as 09$-$18, 17$-$11 and 37$-$11, following the nomenclature used in \citep[][see also §~\ref{subsubsection: Initial Conditions}]{libeskind2020hestia}. We show the \emph{LG distance} (defined as the distance of a galaxy from the geometric centre of the line that connects MW and M31), the mass in stars and gas bound to each galaxy, and $R_\text{200}$ and $M_\text{200}$ of each galaxy. SFR is the star formation rate for all the gas cells within twice the stellar half mass radius. $Z_\text{SFR}/Z_{\odot}$ is the SFR-weighted gas metallicity, normalized with respect to the solar metallicity. We also list the observational estimates for MW from \citet{bland2016} as well as that for M31. The observational estimate for $R_\text{200}$ of the MW is calculated as $R_\text{200} = 1.3 \times R_\text{vir}$ (following \citealt{van2012m31}) from the $R_\text{vir}$ value for MW given in \citet{bland2016}.}
\label{table:Relevant parameters for the three systems}
\begin{minipage}{\textwidth} \centering
\begin{tabular}{c|rr|rr|rr|c|c}
\hline
\hline
& \multicolumn{2}{c|}{09$-$18} & \multicolumn{2}{c|}{17$-$11} & \multicolumn{2}{c|}{37$-$11}&
Obs. estimates for MW & Obs. estimates for M31\\
& MW & M31 & MW & M31 & MW & M31 & (from \citealt{bland2016})\\
\hline
\hline
{LG distance (kpc)} & 433.19 & 433.19 & 338.01 & 338.01 & 425.29 & 425.29 & - & -\\
{log $M_*$ (M$_{\odot}$)} & 10.91 & 11.11 & 11.06 & 11.08 & 10.77 & 10.72 & 10.69 $\pm$ 0.088 & 10.84-11.10 \footnote{\citet{sick2015stellar,rahmani10.1093/mnras/stv2951}}\\
{log $M_\text{gas}$ (M$_{\odot}$)} & 11.08 & 11.20 & 10.92 & 11.21 & 10.76 & 10.87 & 10.92 $\pm$ 0.067 & 9.78 \footnote{\citet{YinrefId0}}\\
{log $M_\text{200}$ (M$_{\odot}$)} & 12.29 & 12.33 & 12.30 & 12.36 & 12.00 & 12.013 & 12.05 $\pm$ 0.096 & 12.10 \footnote{\label{lehner2020}\citet{lehner2020project}}\\
{$R_\text{200}$ (kpc)} & 262.54 & 270.40 & 264.48 & 277.46 & 211.25 & 212.81 & 216.93 $\pm$ 23.075 & 230 \footnotemark[3]{}\\
{log $M_\text{BH}$ (M$_{\odot}$)} & 8.1314 & 8.277 & 7.7362 & 8.1838 & 7.7787 & 7.5840 & 6.6232 $\pm$ 0.02 & 8.15 \footnote{\citet{schiavi2020}}\\
{SFR (M$_{\odot}$ yr$^{-1}$)} & 9.476 & 3.757 & 3.600 & 4.754 & 1.193 & 2.337 & 1.650 $\pm$ 0.19 & 0.25-1.0 \footnote{\citet{williams2003recent,williams2003MNRAS.340..143W,barmby2006dusty,tabatabaei2010,ford2013,lewis2015ApJ...805..183L,rahmani10.1093/mnras/stv2951,boardman2020}}\\
{$M_*$/SFR (Gyr)} & 8.577 & 34.29 & 31.89 & 25.29 & 49.36 & 22.46 & - & -\\
%{log ($M_*$/SFR) Gyr} & 1.1 & 0.5 & 0.56 & 0.47 & 0.08 & 0.4 & \numrange{-0.3}{0.5}\\
$Z_\text{SFR}/Z_{\odot}$ & 3.26 & 3.04 & 3.55 & 3.05 & 3.26 & 3.31 & - & -\\
\hline
\hline
\end{tabular}
\end{minipage}
\end{table*}

\begin{table*}
\caption{For the ions considered in our ionization analysis, we list the wavelength of the strongest tracer ion transitions, the ionization energy, characteristic temperature and characteristic density (we quote the ionization energy values from \citealt{Edl_n_1979,Martin1983,JOHNSON1985405,drake1988,jentschura2005energy}-- obtained from the NIST Database; remaining values are quoted from the supplemental fig. 4 in \citealt{tumlinson2017}). Ionization energy is the energy required to ionize a species into its corresponding higher ion state (in this case, each of the five ions included in our analysis). Our ionization modelling is carried out with {\sc cloudy}v17.}
\label{table:Ion params}
\centering
\begin{tabular}{crrcc}
\hline
\hline
Ion & Wavelength (\si{\angstrom})

& Ioniz. energy (eV) & $\log (T/\text{K})$ & $\log (n_\text{H}$/cm$^{-3}$)\\
\hline
\hline
H{\,\sc i} & 1216.00 & 13.6 & 4.0-4.5 & $\sim$2.0\\
Si{\,\sc iii} & 1206.00 & 33.5 & <5.0 & $-2.5$\\
O{\,\sc vi} & 1031.00 & 138.12 & 5.5 & $-4.5$\\
O{\,\sc vii} & 21.00 & 739.29 & 5.9 & $-5.0$\\
O{\,\sc viii} & 18.96 & 871.41 & 6.4 & $-5.5$\\
\hline
\hline
\end{tabular}

\end{table*}

%10.19,10.28,113.90,138.12,739.33

The moving-mesh magneto-hydrodynamic code, {\sc arepo} \citep{springel2010pur, pakmor2016improving}, has been employed for the higher resolution runs. {\sc arepo}, which is based on a quasi-Lagrangian approach, uses an underlying Voronoi mesh (in order to solve the ideal MHD equations) that is allowed to move along the fluid flow, thus seamlessly combining both Lagrangian as well as Eulerian features in a single cosmological simulation. The code follows the evolution of magnetic fields with the ideal MHD approximation \citep{pakmor2011, pakmor2013} that has been shown to reproduce several observed properties of magnetic fields in galaxies \citep{Pakmor_2017, pakmor2018} and the CGM \citep{Pakmor_2020}. Cells are split (i.e. refined) or merged (i.e. de-refined) whenever the mass of a particular mesh cell varies by more than a factor of two from the target mass resolution. 

We adopt the Auriga galaxy formation model \citep{grand2017auriga}. A two-phase model is used to describe the interstellar medium (ISM), wherein a fraction of cold gas and a hot ambient phase is assigned to each star-forming gas cell \citep{springel2003cosmological}. This two-phase model is enabled for gas denser than the star formation threshold ($0.13$ cm$^{-3}$). Energy is transferred between the two phases by radiative cooling and supernova evaporation, and the gas is assumed to be in pressure equilibrium following an effective equation of state (similar to fig.~4 in \citealt{2005MNRAS.361..776S}). Stellar population particles are formed stochastically from star-forming cells. Black holes (BH) formation and their subsequent feedback contributions are also included in the Auriga framework. Magnetic fields are included as uniform seed fields at the beginning of the simulation runs ($z = 127$) with a comoving field strength of $10^{-14}$ G, which are amplified by an efficient turbulent dynamo at high redshifts \citep{Pakmor_2017}. Gas cooling via primordial and metal cooling \citep{vogelsberger2013model} and a spatially uniform UV background \citep{faucher2009} are included. Our galaxy formation model produces a magnetized CGM with a magnetic energy, which is an order of magnitude below the equipartition value for the thermal and turbulent energy density \citep{Pakmor_2020}.

In our galaxy formation model, the CGM experiences heating primarily from sources such as SNe Type II, AGN feedback (see fig.~17 in \citealt{grand2017auriga}), stellar winds and time-dependent spatially uniform UV background. Stellar and AGN feedback are especially important since they heat and deposit a substantial amount of metals as well as some baryonic material into the CGM \citep{vogelsberger2013model, Bogd_n_2013}.

We do not include extra-planar type Ia SNe or runaway type II SNe. We expect the uncertainty due to not including these in our physics model to be extremely small with respect to that due to treating the ISM with an effective equation of state \citep[see for example fig. 10 in ][]{2019MNRAS.489.4233M}.

Quasar mode feedback is known to suppress star formation in the inner disc of galaxies (particularly relevant at early times) while the radio mode feedback is known to control the ability of halo gas to cool down efficiently at late times (hence relevant in the context of this study). In general, radio mode feedback is instrumental in keeping the halo gas hot, which in turn results in lesser cool gas in the CGM (see fig.~17 from \citealt{grand2017auriga}; also fig.~9 from \citealt{irodotou2021effects}). \citealt{Hani_2019} studied the effect of AGN feedback on the ionization structure within the CGM of a sample of MW-like galaxies from the Auriga simulations. On the whole, they concluded that in comparison to the galaxies without any AGN feedback, the CGMs of galaxies with an AGN feedback exhibited lesser column densities for low and intermediate ions while the column densities for high ions remained largely unchanged. While the presence of an ionizing AGN radiation field in the CGM is responsible for slightly reduced abundances of low ions, the abundances of high ions like O{\,\sc vi} mainly arise from the halo virial temperatures and are hence, largely unaffected by the AGN feedback effects.

We use the {\sc subfind} halo finder \citep{springel2001, dolag2009,2020arXiv201003567S} to identify galaxies and galaxy groups in our analysis. When the simulations were run, black holes were seeded in haloes identified by {\sc subfind}. 

Our simulations and analysis consistently use the Planck 2014 best-fit cosmological parameters \citep{ade2014planck}, which have the following values: $H_{\text{0}} = 100h$ km s$^{-1}$ Mpc$^{-1}$, where $h = 0.677$, $\sigma_{8} = 0.83$, $\Omega_{\Lambda} = 0.682$, $\Omega_\text{M} = 0.270$ and $\Omega_\text{b} = 0.048$.

\subsubsection{Global properties of the \hestia analogues}
\label{subsubsection:global}

Table~\ref{table:Relevant parameters for the three systems} lists key properties for the three realizations of the MW-M31 analogues, which we consider in this paper. We define $R_{200}$ as the radius within which the spherically averaged density is 200 times the critical density of the universe. $M_{200}$ is the total mass within $R_{200}$. The overall $M_{200}$, $M_{*}$, $M_\text{gas}$ and $R_{200}$ values for our MW-M31 analogues are broadly consistent with typical observational estimates (see fig.~7 in \citealt{libeskind2020hestia}; see also \citealt{bland2016,YinrefId0}). Among the two most massive galaxies in each of our LG simulations, the galaxy with a larger value of $M_{200}$ is identified as M31, while the other galaxy is identified as MW.

The \hestia MW analogues reveal $M_\text{BH}$ values an order of magnitude larger than that stated in the observations of \citet{bland2016}. This does not, however, necessarily mean that the AGN feedback has been too strong during the simulations, because we see realistic MW stellar masses at $z=0$. For the CGM, which we study extensively in this paper, the overestimated MW BH masses therefore do not necessarily indicate too strong AGN feedback. We also note that our MW analogues are still consistent with MW-mass galaxies (see fig.~5 of \citealt{2016ApJ...817...21S}).

Similarly, the SFR at $z=0$ is also comparable to or larger than observed. We note that the SFR of M31 is larger by a factor of a few in \hestia in comparison to observations. The generation of winds is closely tied to the SFR in our simulations, so it is possible that the role of outflows is over-estimated by \hestia in comparison to the $z=0$ observations of M31. Integrated over the lifetime of the galaxies, \hestia does, however, produce realistic stellar masses at $z=0$\footnote{At a speculative node, it is possible that a too high SFR could be compensated by too high AGN feedback, and this would result in a stellar mass consistent with observations but at the same time a too massive BH mass. Addressing such a hypothesis would require running additional simulations which is beyond the scope of this paper.}. We, therefore, do not regard the discrepancy between the $z=0$ SFR as more problematic than the uncertainty already in place by using an effective model of winds, or, for example, by the simulated M31 galaxies having different merger histories or disc orientations than the \emph{real} M31. In comparison to the SFR values from \citealt{bland2016}, other observational studies report slightly larger SFR values for MW (1-3 M$_{\odot}$ yr$^{-1}$, 3-6 M$_{\odot}$ yr$^{-1}$, 1-3 M$_{\odot}$ yr$^{-1}$, 1.9 $\pm$ 0.4 M$_{\odot}$ yr$^{-1}$: \citealt{mckee1997ApJ...476..144M,boissier10.1046/j.1365-8711.1999.02699.x,wakker2007ApJ...670L.113W, wakker2008ApJ...672..298W,chomiuk2011toward}), but these are nevertheless lower than the \hestia values. We also notice that the MW analogue in the 09$-$18 simulation exhibits a substantially higher SFR than others. However, all $M_*/$SFR values (except those for the MW analogue in 09$-$18 simulation) for our sample are still well within the observational constraints of normal star-forming galaxies with masses comparable to the MW and M31 (see fig.~8 in \citealt{speagle2014ApJS..214...15S}). Thus, overall, the \hestia galaxies seem to be slightly more star-forming in comparison to the observations but this does not induce larger uncertainties in our analysis than already present due to multiple other factors which we highlighted earlier.

We also note that the SFR-averaged gas metallicity is consistent with the M31 measurement in \citet{sanders2012ApJ...758..133S}.

\subsection{Analysis}
\label{subsection: Analytical methods}

In this section, we describe the methodology adopted in order to compute the ion fractions in the CGM, underlying assumptions, their possible effects on the interpretation of our results and the process of creating Mollweide maps from the computed ion fractions. We make use of the photo-ionization code \textsc{cloudy} to obtain ionization fractions for the tracer ions H{\,\sc i}, Si{\,\sc iii}, O{\,\sc vi}, O{\,\sc vii} and O{\,\sc viii}, and we generate Mollweide projection maps using the \textsc{healpy} package to create mock observations.

\subsubsection{Ionization modelling with {\sc cloudy}}
\label{subsubsection: Ionization}

Two principal ionization processes in the CGM and IGM are collisional ionization and photo-ionization \citep{bergeron1986absorption, prochaska2004host,turner2015detection}. An equilibrium scenario is generally assumed in both these processes thus resulting in a collisional ionization equilibrium (CIE) and a photo-ionization equilibrium (PIE). 

Such a bi-modal attempt in the ionization modelling has to date proved to be sufficient to well explain the co-habitation of both high and low ions in different phases at the same time within a common astrophysical gas environment. Generally, high ions (e.g., O{\,\sc vii}, O{\,\sc viii}, Ne{\,\sc viii}, Mg{\,\sc x}) are found to be better modelled via CIE while the low and intermediate ions (e.g., Fe{\,\sc ii}, N{\,\sc i}, S{\,\sc ii}) lend themselves better to PIE, owing to the temperatures in the various gas phases and the strength and shape of the UV background field. 
CIE, which assumes that the ionization is mainly carried by \text{electrons}, can be well characterised \citep{richter2008} using the relation,
\begin{equation}
\label{equation: CIE}
f_\textmd{H{\,\sc i},coll} = \frac{\alpha_\textmd{{H}}(T)}{\beta_\textmd{{H}}(T)},
\end{equation}
where $f_{\textmd{H{\,\sc i},coll}}$ is the neutral hydrogen fraction in CIE, $\alpha_{\textmd{H}}(T)$ is the temperature dependent recombination rate of hydrogen and $\beta_{\textmd{H}}(T)$ is the collisional ionization coefficient, both for hydrogen. 

PIE, on the other hand, assumes \text{photons} to be the primary perpetrators and can be better described \citep{richter2008} as,
\begin{equation}
\label{equation: PIE}
f_{\textmd{H{\,\sc i},photo}} = \frac{n\textsubscript{e}   \alpha_\textmd{H}(T)}{\Gamma_\textmd{H{\,\sc i}}},
\end{equation}
where $f_{\textmd{H{\,\sc i},photo}}$ is the neutral hydrogen fraction in PIE, $n\textsubscript{e}$ is the electron density and $\Gamma_\textmd{H{\,\sc i}}$ is the photo-ionization rate.

We determine the ionization fractions using the {\sc cloudy} code (version C17; \citealt{Ferland2017}), which is designed to model photo-ionization and photo-dissociation processes by including a wide combination of temperature-density phases for a list of elements, in order to simulate complex astrophysical environments realistically and produce mock parameters and outputs. The temperature of each {\sc arepo} gas cell is given as input to {\sc cloudy} (in practice, we use lookup tables to speed up the calculation, see below), which determines the ionization state in post-processing. For the star-forming gas cell we directly set all atoms to be neutral, because most of the mass is in the cold phase.

We include both CIE and PIE in the modelling code. The UV background from \cite{faucher2009} is used. Self-shielding prescriptions, in particular for H{\,\sc i} gas in denser regions, are adopted from \cite{rahmati2013impact}. We do not include AGN continuum radiation for the sake of simplicity. While excluding the AGN radiation might affect the ion fractions in regions close to the galaxy (e.g. ISM), it is much less likely to have any dominant impact in the CGM. Our {\sc cloudy} modelling is identical to that introduced in \citet{hani2018}, with the only difference that we use a finer resolution grid for the output tables. In our analysis, we impose a metallicity floor of $10^{-4.5}$ $Z_{\odot}$ to avoid metallicity values lower than those present in our {\sc cloudy} tables. Note that we do not include photo-ionization from stars or AGN in this work.

For this paper, we focus on the five tracer ions listed in Table~\ref{table:Ion params} for which we generate mock observables; two of which are largely representative of the cold and cool-ionized ($T\sim 10^{4}-10^{5}$ K) gas (H{\,\sc i} and Si{\,\sc iii}) and the three ions representative of the warm-hot ($T>10^{5.5}$ K) gas (O{\,\sc vi}, O{\,\sc vii} and O{\,\sc viii}). These five ions have a host of robust corresponding observational CGM data as well (e.g. \citealt{liangchen10.1093/mnras/stu1901, werk2014cos, johnson2015possible, Richter_2016, Richter_2017, lehner2020project}). 

Si{\,\sc iii} may also be produced by photoionization at a much lower temperature than $10^5$ K. However, neither does our ionization modeling include photoionization from stars nor is it optimal in describing gas colder than $10^4$ K. Therefore, this remains an uncertainty in our ionization modelling.

The overall ion abundances are naturally depending on the gas metallicity distribution in \hestia. In Appendix~\ref{RadialMetallicity} we, therefore, derive radial gas metallicity profiles for the simulated MW and M31 galaxies (see Fig.~\ref{fig:Metallicity}). We conclude that the disc gas metallicity in \hestia is up to 3 times higher than realistic MW- and M31-mass galaxies \citep{sanders2012ApJ...758..133S, 2014MNRAS.438.1985T}. The gas metallicity profile of the CGM of MW and M31 is not well constrained observationally, but we speculate the \hestia might as well have a slightly too high gas metallicity there. We will keep this in mind when comparing our simulations to observations (see Sec.~\ref{MetallicityNorm}).

\begin{figure*}
\centering
\begin{minipage}{.33\linewidth}
\includegraphics[width=1.0\linewidth]{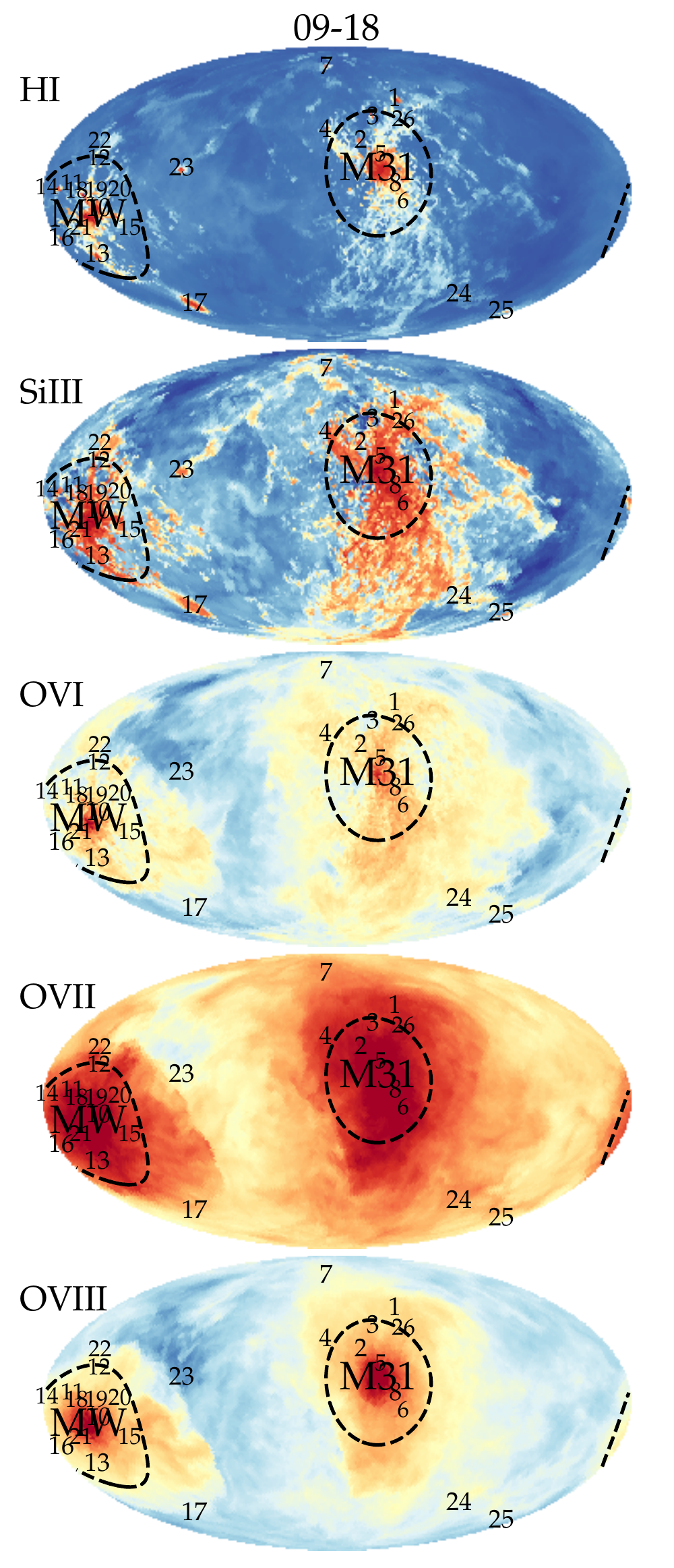}
\end{minipage}
\begin{minipage}{.33\linewidth}
\includegraphics[width=1.0\linewidth]{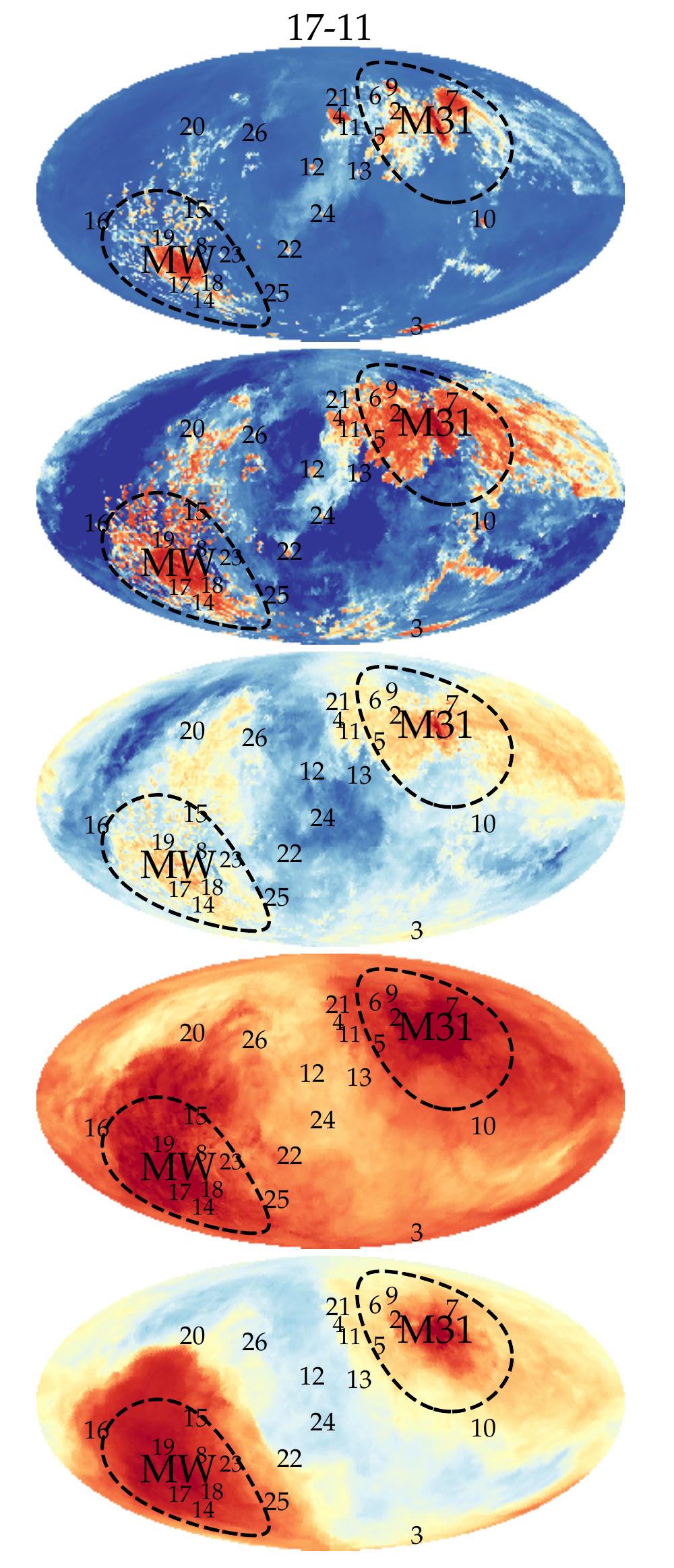}
\end{minipage}
\begin{minipage}{.33\linewidth}
\includegraphics[width=1.0\linewidth]{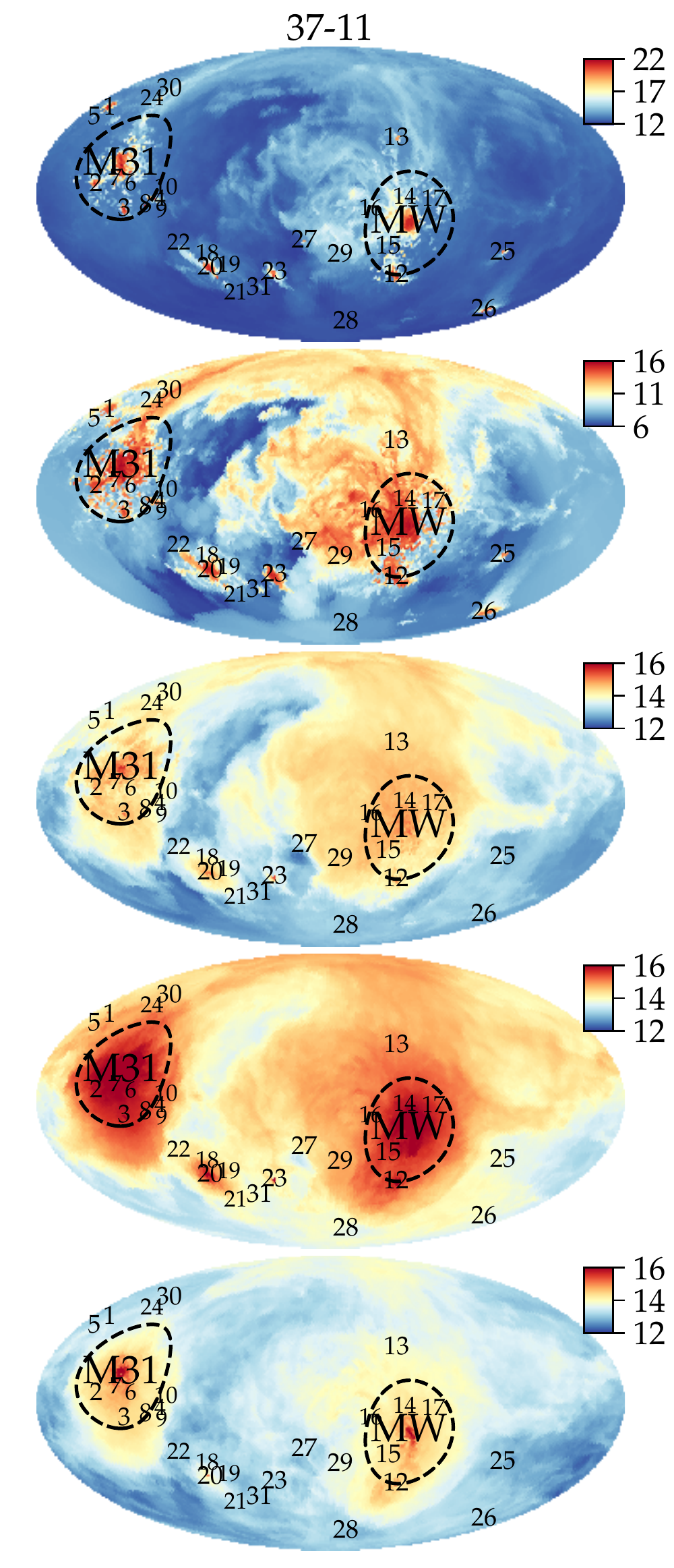}
\end{minipage}
\caption{Skymaps showing the gas column densities in our three LG realizations. The maps are generated from an observer located in the LG centre, so we can see the predictions for both MW and M31 simultaneously. H{\,\sc i} and Si{\,\sc iii} trace cold-dense gas in and around the ISM, whereas O{\,\sc vi}, O{\,\sc vii} and O{\,\sc viii} trace gradually more hot-diffuse halo gas. Colour-bars show the log column densities, \textit{N}, for each ion (\textit{N} in units of cm$^{-2})$. Dashed lines indicate $R_{200}$ of MW and M31. Small numbers indicate the location of galaxies other than M31 and MW. All dense H{\,\sc i} absorbers with $N_{\rm HI}>10^{20}$ cm$^{-2}$ are associated with a galaxy. The distributions of the oxygen ions tracing warmer gas reveal a less clumpy and more spherical distribution around the massive galaxies.
}
\label{Fig:Skymaps}
\end{figure*}

\begin{figure*}
\centering
\begin{minipage}{.33\linewidth}
\centering
\includegraphics[height=5.9cm]{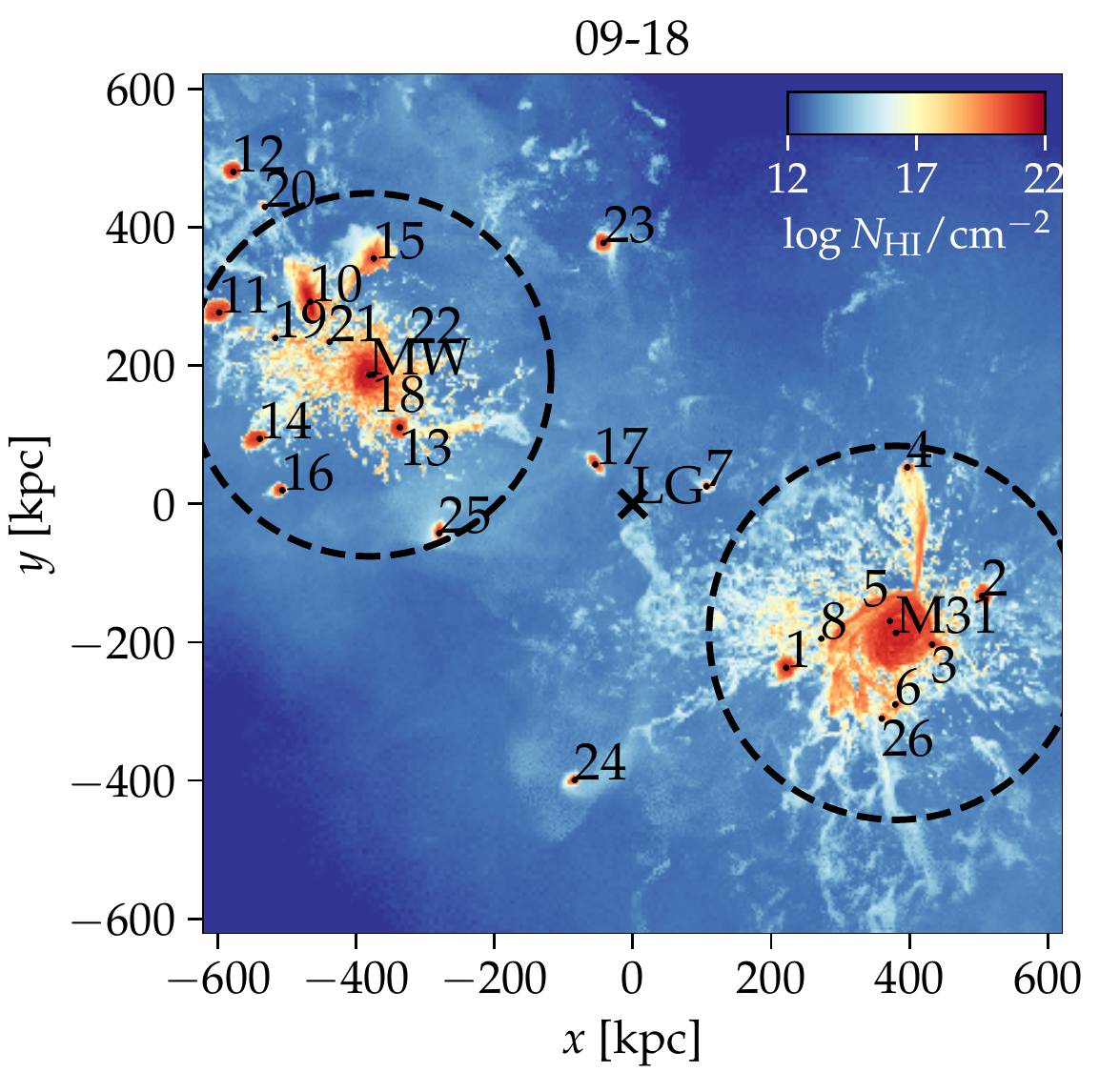}
\end{minipage}
\begin{minipage}{.33\linewidth}
\centering
\includegraphics[height=5.9cm]{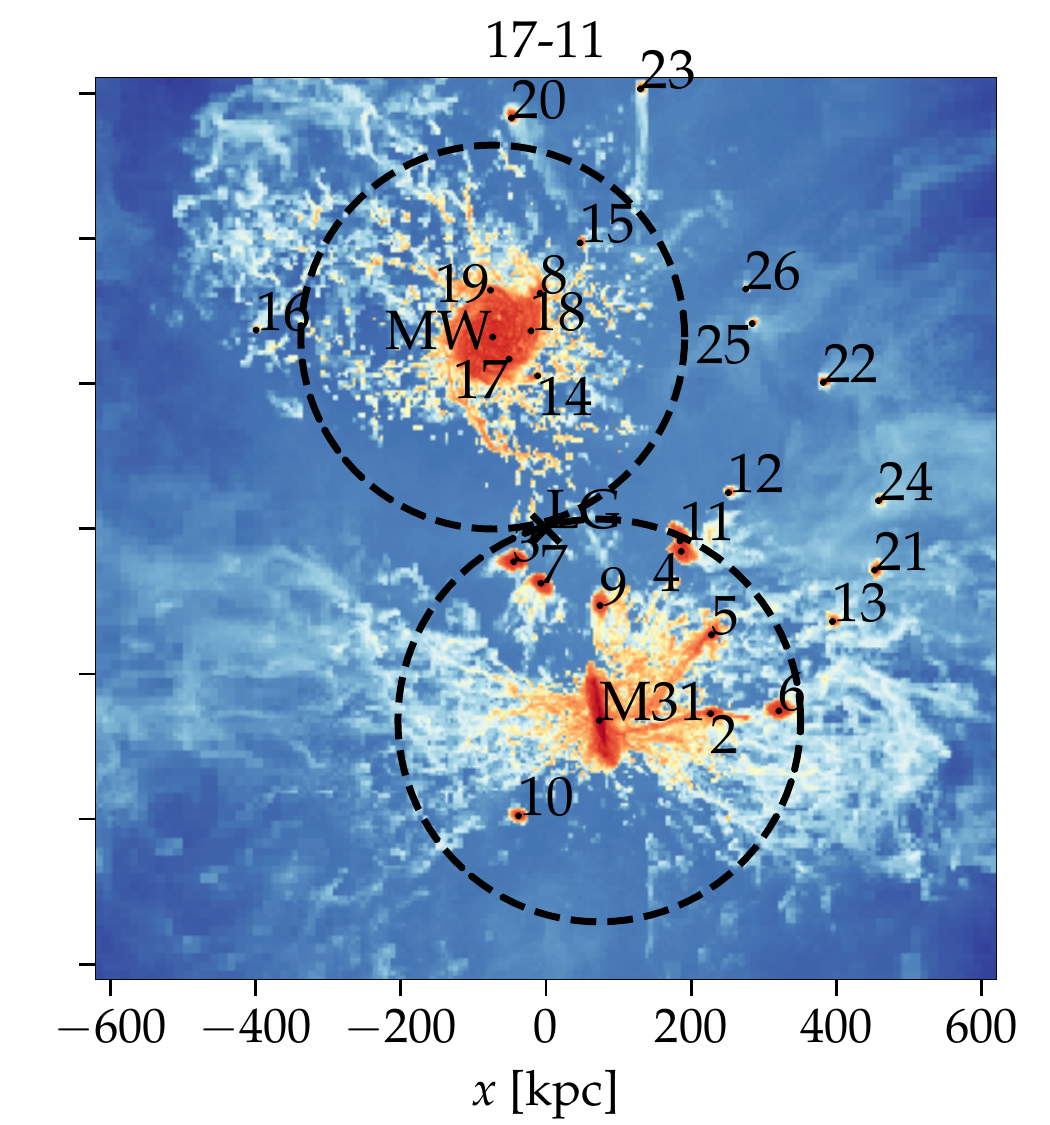}
\end{minipage}
\begin{minipage}{.33\linewidth}
\centering
\includegraphics[height=5.9cm]{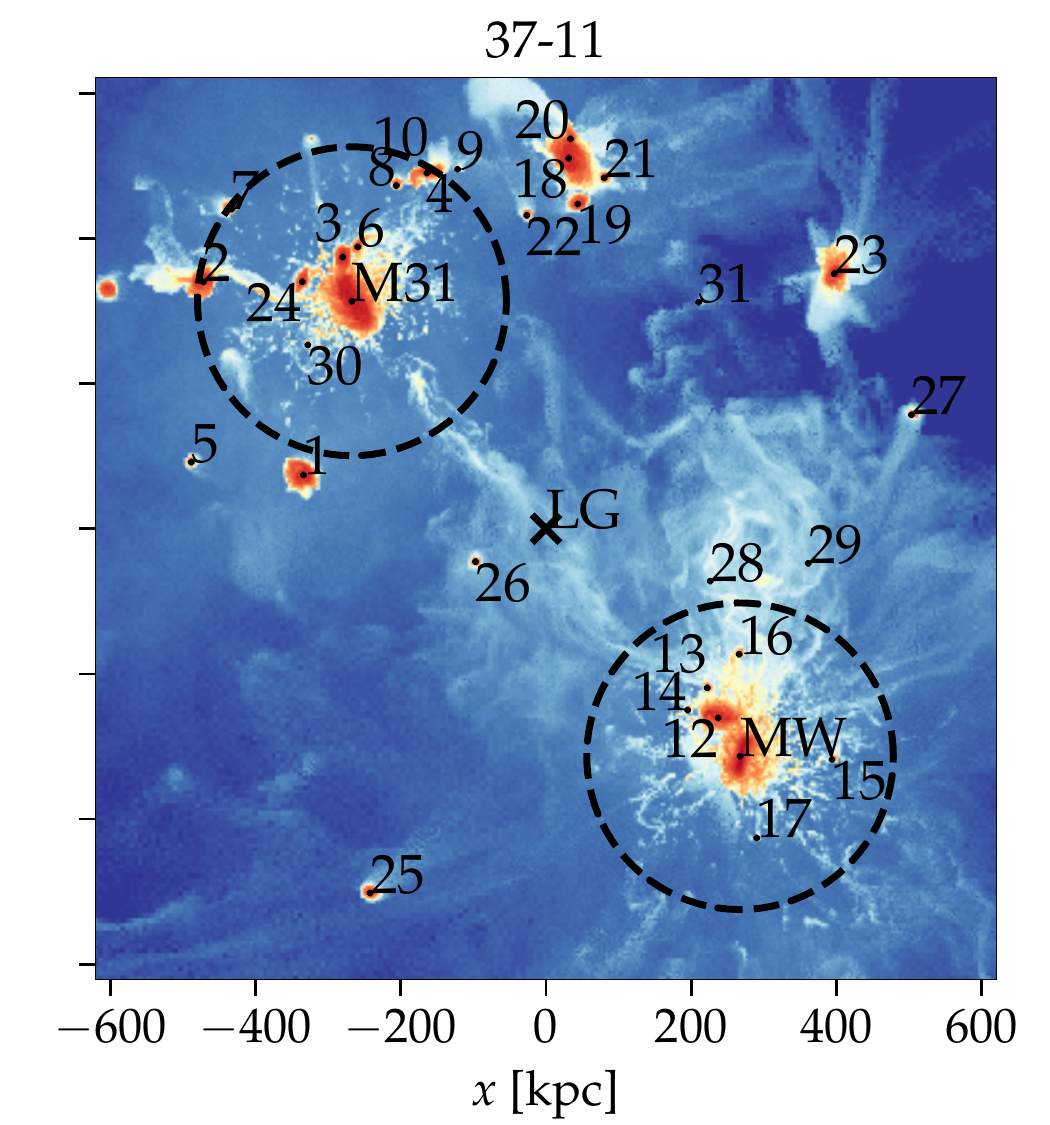}
\end{minipage}
\caption{H{\,\sc i} column density projection maps for each realization. The colour-bar denotes the range of log H{\,\sc i} column densities (\textit{N} in units of cm$^{-2}$). The dashed circles indicate $R_{200}$ of M31 and MW. Note that the actual spatial extent spanned by galaxy number 7 in the 17$-$11 realization is far smaller as compared to its projected spatial extent in the corresponding skymap (Fig.~\ref{Fig:Skymaps}; see also §~\ref{cartesian}).}
\label{Fig>Projections}
\end{figure*}

\subsubsection{Generation of skymaps}
\label{subsubsection: healpy}

The analysis in this paper extensively uses \emph{skymaps} showing the column density distribution of the different ions. To define the unit vectors characterising each sightline, we use the Mollweide projection functionality from the {\sc healpy} package \citep{Zonca2019}, associated to the {\sc HEALPix}-scheme \citep{Gorski2005}. Each {\sc HEALPix} sphere consists of a set of pixels (12 pixels in three rings around the poles and equator) that give rise to a \textit{base resolution}. The grid resolution, $N_{\rm side}$, denotes the number of divisions along the side of each base-resolution pixel. The total number of equal-area ($\Omega_{\rm pix}$) pixels, $N_{\rm pix}$, can be expressed as, $N_{\rm pix} = 12  N_{\rm side}^2$. The area of each pixel is, $\Omega_{\rm pix}= \pi/(3N_{\rm side}^2)$ and the angular resolution per pixel is, $\Theta_{\rm pix} = \Omega_{\rm pix}^{1/2}$. 
We select $N_{\rm side} = 40$ for all the Mollweide projection plots in this paper. This yields the total number of pixels (which we hereafter refer to as \textit{sightlines}) $N_{\rm pix} = 19,200$, and an angular resolution $\Theta_{\rm pix} = 1.46^{\circ}$.

Each sightline starts in $(x,y,z)=(0,0,0)$ (we shift our coordinates to our desired origin; see §~\ref{section: Results} for further discussion) and ends 700 kpc away in the direction of the unit vector defined by the HEALPix-pixel. A sightline is binned into 50,000 evenly spaced gridpoints, so we get a grid-size of 14 pc. At each gridpoint we set the ion density equal to the value of the nearest gas cell (we use the {\sc scipy}-function, KDTree, to determine the nearest neighbour). The projected ion column density for a sightline is then calculated by summing the respective ion number densities over the grid-points constituting a sightline.

\section{Results}
\label{section: Results}

\subsection{Skymaps}
\label{Skymaps}

We create skymaps centred on the geometric centre of the LG, which we define to be the midpoint between MW and M31. Based on such skymaps we will later compute projected column density profiles of M31 (see §~\ref{ColDens}), which makes it possible to directly compare our simulations to observations. A similar frame-of-reference also proved useful for \citet{nuza2014distribution}; though they used it to obtain plots for studying the entire LG but did not produce whole skymaps from this point.

Fig.~\ref{Fig:Skymaps} shows Mollweide projections of the skymaps for each of the five ions -- H{\,\sc i}, Si{\,\sc iii}, O{\,\sc vi}, O{\,\sc vii} and O{\,\sc viii} -- for each LG realization (Table~\ref{table:Relevant parameters for the three systems}). All ions reveal an over-density centred on the MW and M31. In this order, the ions trace gradually warmer gas, and it is, therefore, not surprising that we see a gradually more diffuse distribution in the projection maps. H{\,\sc i} and Si{\,\sc iii} are much more centred on the inner parts of the haloes in comparison to O{\,\sc vi}, O{\,\sc vii} and O{\,\sc viii}, the latter filling the space all way out to $R_{200}$ (and even beyond).  $R_{200}$ is shown as dashed circles around MW and M31 -- note, that a circle in a Mollweide projection appears deformed. 

We see that all dense gas blobs with $N_\text{HI}>10^{20}$ cm$^{-2}$ are associated with regions overlapping with the galaxies from our catalogue (see Appendix \ref{Listing}). It fits well with the expectation that such high column densities are typically associated with the ISM of galaxies. The CGM regions of the MW and M31 analogues show a rich structure of H{\,\sc i}-features. In 09$-$18, the M31 analogue, for example, reveals a bi-conical structure, characteristic of galaxy outflows. Many of the extended, diffuse gas streams (particularly in Si{\,\sc iii}, but also in H{\,\sc i}) go far beyond the haloes of the MW and M31 analogues. We see varying distributions of Si{\,\sc iii} gas across the three simulations in corresponding skymaps. While the 17$-$11 and 09$-$18 simulations show an excess of higher column density, clumpy Si{\,\sc iii}, 37$-$11 shows an excess of lower column density, diffuse Si{\,\sc iii} (See §~\ref{coherence} and \ref{ColDens} for further discussion). Smaller stellar mass and $R_{200}$ values for MW-M31 in case of 37$-$11 (in comparison to the other two simulations) could be one possible reason for such a heterogeneity across the Si{\,\sc iii} distributions.

\subsubsection{Satellite galaxies in the LG}
\label{Satellite galaxies in the LG}

The satellite galaxies in the simulations have been marked with a galaxy number in each panel, and their properties are summarised in the catalogue tables in Appendix~\ref{Listing}. We include those galaxies which have $M_\text{gas}>0$ (as identified by the {\sc subfind} halo finder) and are within 800 kpc of the LG centre. The 800 kpc cutoff is slightly larger in comparison to the 700 kpc cutoff, used when generating the skymaps; we have chosen this slightly larger cutoff for the satellite galaxy catalogue to ensure that all galaxies contributing to the skymaps are included. Below, we show that all dense H{\,\sc i} blobs are associated with a galaxy from our catalogues, so a 800 kpc cutoff sufficiently selects all the galaxies contributing to the skymaps.

The satellite galaxies are generally more prominent in H{\,\sc i} and Si{\,\sc iii} in comparison to the higher ions. Galaxy 12 from the 37$-$11 simulation does, however, reveal significant amounts of O{\,\sc vi}, O{\,\sc vii} and O{\,\sc viii}. This satellite has a stellar mass of $M_*=3.2\times 10^9 M_{\odot}$, which is comparable to the LMC galaxy in the real LG (it has $M_*=3\times 10^9 M_{\odot}$ following \citealt{2016ARA&A..54..363D}). Recently, it has been suggested that the LMC galaxy may have a warm--hot coronal halo \citep{wakker1998, lucchini2020} that is responsible for the presence and spatial extent of the Magellanic Stream.

\citet{adams2013catalog} presents a study of 59 ultra compact high-velocity clouds (UCHVCs) from the ALFALFA H{\,\sc i} survey while \citet{giovanelli2013alfalfa} reports the discovery of a low-mass halo in the form of a UCHVC. From both these studies, a common conclusion emerges: low-mass, gas-rich halos (detected in the form of Compact HVCs/UCHVCs), lurking on the fringes of the CGMs of massive galaxies in our Local Volume (MW-M31, for example), are more likely to be discovered through their baryonic content (traceable primarily via H{\,\sc i}). Other observational papers \citep{deheij2002A&A...391..159D,Putman_2002,Sternberg_2002,Maloney_2003,westmeiera2005A&A...436..101W,westmeierb2005A&A...432..937W}, based on objects detected around MW and M31, also support this hypothesis. We find similar H{\,\sc i} column densities ($\sim 10^{19}$ cm$^{-2}$) at  $R_\text{proj} \lesssim  R_\text{200}$ ($\lesssim 200$ kpc), as reported in the above observations. One can also very clearly notice the presence of such small halos in our skymaps. Hence, we can safely conclude that our results also support the existence of low-mass halos at circumgalactic distances.

\subsubsection{Ram pressure stripping in the LG}
\label{Ram pressure stripping in the LG}

Many of the satellite systems show \emph{disturbed} H{\,\sc i} and Si{\,\sc iii} gas distributions (see the satellite galaxies with galaxy numbers 4 and 13 for simulation 09$-$18; 2, 5 and 7 for simulation 17$-$11; 2 for simulation 37$-$11) to varying degrees. The satellite galaxies' proximity to either of MW or M31 certainly plays a pivotal role (as do their own kinematic motions through their surroundings) in producing ram pressure stripping in their ISMs as well as generating asymmetries in their respective CGMs \citep{simpson2018quenching, hausammann2019A&A...624A..11H}. The ions tracing the warmer gas appear to be less sensitive to such disturbances. 

In the context of galaxy clusters, ram-pressure stripping of the ISM gas is an important process in quenching galaxies \citep{1972ApJ...176....1G, 1999MNRAS.308..947A}. \emph{Jellyfish galaxies} are examples of galaxies experiencing such stripping, where the ram-pressure from intracluster gas strips and disturbs the ISM of star-forming galaxies \citep{2017ApJ...844...48P,2019ApJ...870...63C}. Such galaxies have long extended tails, which are stabilised by radiative cooling and a magnetic field \citep{2020NatAs.tmp..224M}. Given the many disturbed galaxies with extended tails in our simulations, we argue that observations of such galaxies may provide insights into the same processes, which are usually studied in jellyfish galaxies in galaxy clusters. It would specifically be interesting, if such examples of jellyfish galaxies in the LG could be used to provide insights into the growth of dense gas in the galaxy tails. The growth of dense gas in such a multiphase medium has recently been intensively studied in hydrodynamical simulations of a cold cloud interacting with a hot wind \citep{2018MNRAS.480L.111G, Li2019,sparre2020interaction,2021MNRAS.501.1143K,2021arXiv210110344A}. We further discuss the possibility of constraining gas flows in the tails of LG satellites in §~\ref{Discussion}.

\begin{figure*}
\centering
\includegraphics[width = 1.0\linewidth]{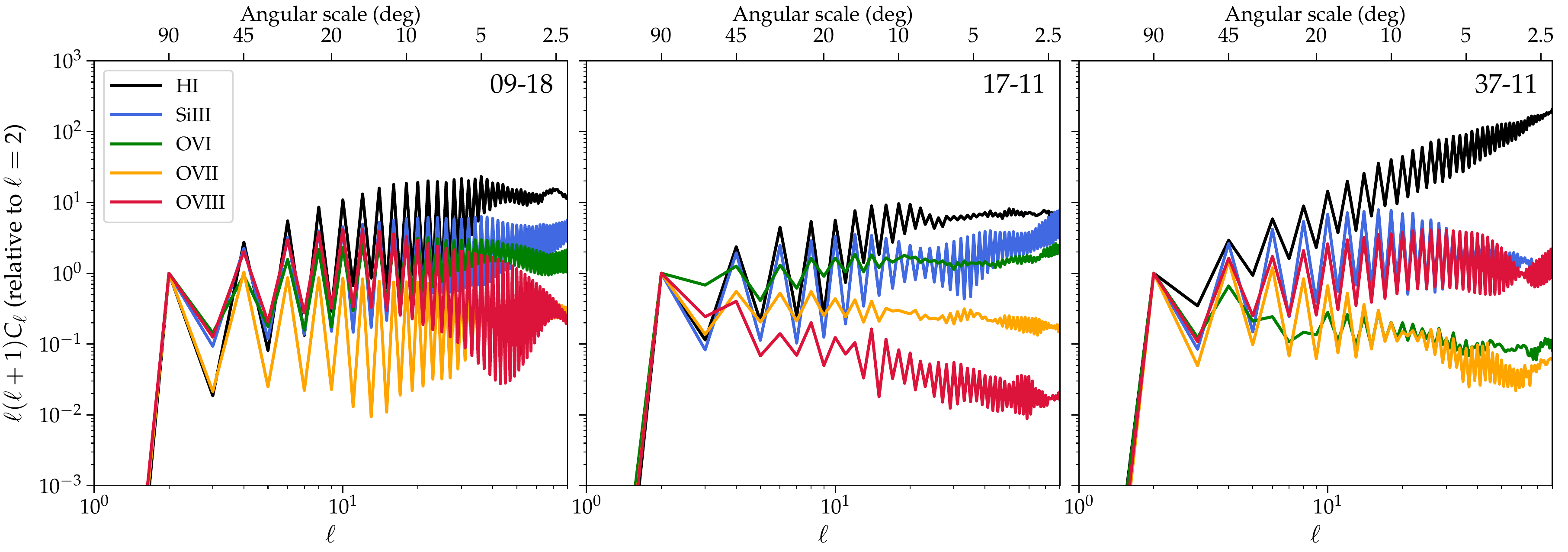}
\caption{We show power spectra generated based on the ion skymaps (Fig.~\ref{Fig:Skymaps}). The power spectra are normalised to the $l=2$ value. The ions tracing the coldest gas (H{\,\sc i} and Si{\,\sc iii}) have more power on small angular scales ($\lesssim 10^\circ$) in comparison to the high ions O{\,\sc vi}, O{\,\sc vii} and O{\,\sc viii}. This fits well with the visual impression from the skymaps in Fig.~\ref{Fig:Skymaps}. The power spectra reveal a preference for modes with even $l$-values. This is because the skymaps have a contribution from a reflective component, with MW and M31 being in opposite directions (as seen from the frame-of-reference of the skymap's observer).}
\label{Power spectra}
\end{figure*}

\begin{figure*}
\centering
\includegraphics[width = \linewidth]{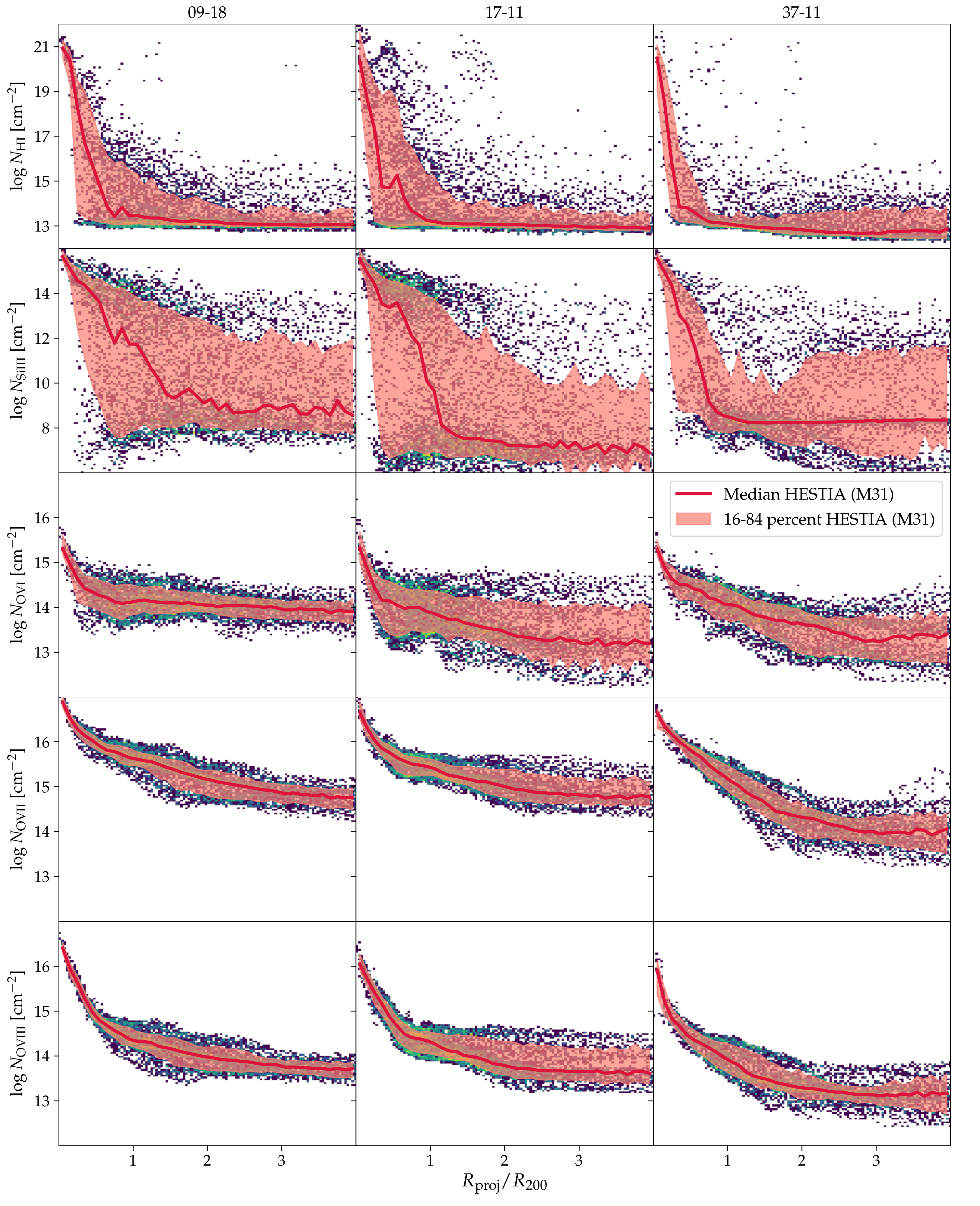}
\caption{\textit{Top-Bottom}: 2D log radial column density profiles for M31 for H{\,\sc i}, Si{\,\sc iii}, O{\,\sc vi}, O{\,\sc vii} and O{\,\sc viii}. Thick, red curve signifies median values while the red, shaded region denotes 16-84th percentile values. The background points depict the ion column density contributions arising from the remaining gas cells. A distinct blob of high column density H{\,\sc i} absorbers, which can be seen at a distance of $\sim$ 1.5 $R_\text{proj}/R_\text{200}$ in the H{\,\sc i} profile for 17$-$11, can be correlated with satellite galaxies numbered 4 and 11 in the corresponding skymap (H{\,\sc i} skymap for 17$-$11 in Fig.~\ref{Fig:Skymaps}).}\label{fig:M31 ColDensSubplt}
\end{figure*}

\subsection{Cartesian projections}
\label{cartesian}
	
The inescapable nature of skymap projections often teases one into a likely misinterpretation of the angular extent spanned by objects within them. This misinterpretation, however, is circumvented by Cartesian projection plots. An example for this is the case of satellite galaxy number 7 in the 17$-$11 realization. Its distance to the LG midpoint is only 114 kpc (see Table~\ref{tab:1711Catalog} in Appendix~\ref{Listing}), which is much smaller in comparison to M31 (338 kpc). In the H{\,\sc i} skymap, this galaxy appears much larger in comparison to the Cartesian projection, which we present in Fig.~\ref{Fig>Projections}. This galaxy, hence, appears to be visually dominant in the Mollweide projection map simply because of its proximity to the LG centre, and its location on the skymap, where it appears to be in the direction of the M31 analogue. Thus, this example demonstrates that it is much harder to distinguish galaxies in the skymap in comparison to a Cartesian projection, which should be kept in mind when it comes to the visual interpretation of the skymaps.

In our analysis, all large and dense H{\,\sc i} blobs with $N_\text{HI}>10^{20}$ cm$^{-2}$ are associated with a galaxy from our galaxy catalogue. There is a minor blob at $x<-600$ kpc in the 37$-$11 H{\,\sc i} projection map (Fig.~\ref{Fig>Projections}), which is not included in the catalogue, because its distance to the LG centre is larger than our cutoff value of 800 kpc -- hence it is not present in the skymaps, but only visible in the Cartesian projection. 

We also see that the satellite galaxies, which we described in §~\ref{Skymaps} as having \emph{disturbed} gas distributions according to the skymaps, also look disturbed in Fig.~\ref{Fig>Projections}. Indeed, the deformed nature seems even more pronounced in the Cartesian projection.

\subsection{Power Spectra}
\label{PowerSpec}

In the previous subsections, we have clearly seen that the low ions largely follow a clumpy distribution while the high ions follow a much smoother profile. One way to neatly quantify such distribution patterns is by creating power spectra for each ion and capture the scales over which the corresponding ion exhibits most of its power.

\subsubsection{Formalism}
\label{formalism}

The spatial scales contributing to a skymap can be quantified by a power spectrum. First, the column density of a given ion is decomposed into spherical harmonics as
\begin{align}
N_\text{ion}(\pmb{r}) = \sum_{lm} a_{lm} Y_l^m (\pmb{r}),
\end{align}
where $\pmb{r}$ is a pixels unit vector, $l$ is the multipole number, and $a_{lm}$ is the coefficient describing the contribution by the mode corresponding to a spherical harmonics base function ($Y_l^m$). The angular power spectrum is then defined as,
\begin{align}
\label{equation: Power spec}
C_l = \frac{1}{2l+1} \sum\limits_{m} |a_{lm}|^2.
\end{align}
We use the {\sc healpy} function \emph{anafast} to compute $C_l$ for each of the column density skymaps. We have subtracted the monopole and dipole moments, and we constrain the power spectrum to $l\leq 2 N_\text{side}=80$, because contributions at higher $l$ may be dominated by noise (following the {\sc healpix} documentation for the anafast function).

In Fig.~\ref{Power spectra}, we show the power spectra for the different ions. We show the power relative to $l=2$, which makes the $l$-dependence for the different ions easy to compare. We have scaled the $C_l$ by a factor of $l(l+1)$, so the plot shows the total power contributed by each multipole. The angular scale corresponding to each multipole number is estimated as $180^\circ /l$.

\subsubsection{Contributions from odd and even modes}
\label{oddevenmodes}

We start by characterising the 09$-$18 simulation. The modes with even $l$-values are systematically larger in comparison to the modes with odd $l$-values. This \emph{zigzagging} could easily be misinterpreted as an effect of noise, but we remark that it has a physical origin caused by the MW and M31 having a similar angular extent, a similar column density and being located in opposite directions (as seen from the skymap-observer's position). These two galaxies, hence, contribute with an approximate reflection-symmetric signal. Due to the identity, $Y_l^m(-\pmb{r}) = (-1)^l Y_l^m(\pmb{r})$, only the modes with even $l$ contribute to a reflection-symmetric map, so this explains the domination of even modes.

A domination of even modes is especially visible for $l\lesssim 10$ for all ions in all three simulations. For 09$-$18 the domination is also present for higher $l$ for all ions, but for 17$-$11, the signal vanishes at $l\gtrsim 10$ for O{\,\sc vi} and O{\,\sc vii}.

\subsubsection{The angular coherence scale}
\label{coherence}

From the behaviour of the power spectra for 09$-$18, we see that the H{\,\sc i} skymaps have more structure on small scales of $\simeq 5^\circ$ (relative to a larger scale of $l=2$) in comparison to the other ions. The amount of power on this angular scale ($\simeq 5^\circ$) is indeed gradually decreasing from H{\,\sc i}, Si{\,\sc iii}, O{\,\sc vi}, O{\,\sc vii} to O{\,\sc viii} (with the only exception being O{\,\sc viii} in 37$-$11, which shows higher power on this scale than O{\,\sc vi} and O{\,\sc vii}). This is completely consistent with the picture that we get from visually examining the different skymaps in Fig.~\ref{Fig:Skymaps}, where the ions tracing the coldest gas also seem to have the clumpiest distribution on small angular scales.

The behaviour of the power spectra for 17$-$11 and 37$-$11 are broadly consistent with this picture. H{\,\sc i} has more power at smaller scales ($l\gtrsim 20$) across all simulations in comparison to the other four ions. For 37$-$11, O{\,\sc viii} shows more power on small scales in comparison to O{\,\sc vi} and O{\,\sc vii}, which is most likely an effect of the O{\,\sc viii} ion being influenced by outflows (this ion, for example, reveals a bi-conical outflow for MW for 37$-$11 in Fig.~\ref{Fig:Skymaps}).

For 37$-$11, the H{\,\sc i} spectrum reveals the most power on small angular scales -- this fits well with our scenario that H{\,\sc i} gas is clumpy on small scales. For the warmer ions such as O{\,\sc vii} the power is a decreasing function of $l$ (if we ignore the fluctuations caused by even modes having more power in comparison to odd modes), implying that fluctuations on large angular scales are dominating. Similar trends are found in the other simulations. 

Intriguingly, the Si{\,\sc iii} power spectra for 09$-$18 and 17$-$11 show an \textit{increasing} trend at small scales ($\lesssim 10^\circ$), while 37$-$11 Si{\,\sc iii} power spectra shows a \textit{decreasing} trend at similar scales. This pattern is, indeed, coherent with our observations regarding the Si{\,\sc iii} skymaps (see §~\ref{Skymaps}). We discuss this aspect a bit further in §~\ref{ColDens}, where we introduce the column density distributions.

\subsection{Column density profiles}
\label{ColDens}

Radial column density profiles are often used as an observational probe of the spatial distribution of the CGM in galaxies. In Fig.~\ref{fig:M31 ColDensSubplt}, we show the M31 radial profiles for our ensemble of ions with a particular focus on the median and 16-84th percentile of the distributions. The background points show all our sightlines.

\subsubsection{Overall trends}
\label{Overall trends}

As expected, the median column density is a declining function of radius for all ions in all simulations. The scatter is, however, behaving differently. Ions tracing the warm--hot gas (O{\,\sc vi}, O{\,\sc vii} and O{\,\sc viii}) have a much lower scatter in comparison to the ions characteristic of dense--cold gas (H{\,\sc i} and Si{\,\sc iii}). The profiles of the former ions are \emph{well-behaved} and the column density profiles can be well-described as a monotonic decreasing function of projected distance (this feature is well documented for O{\,\sc vi}; \citealt{werk2013,liang2016}) with a scatter of 0.1-0.2 dex. On the other hand, H{\,\sc i} and Si{\,\sc iii} reveal extreme outliers. In simulation 17$-$11 galaxies 4, 11 and 21 (see Fig.~\ref{Fig:Skymaps}), for example, contribute with high H{\,\sc i} column densities ($\gtrsim 10^{20}$ cm$^{-2}$) at a projected radius of $R_\text{proj}=1.5\pm 0.5 R_\text{200}$. This shows that the H{\,\sc i} column density is clumpy and influenced by satellite galaxies. 

Similarly, Si{\,\sc iii} show multiple clumps, but their correlation scales seem slightly larger in comparison to H{\,\sc i}, which is consistent with our power spectrum analysis. Despite the clumpy nature of Si{\,\sc iii}, we still find the mean of the projected column density profile to be decreasing (as, for example, is also seen in the observed sample of galaxies from \cite{liangchen10.1093/mnras/stu1901}).

These trends are also applicable to the projected column density profiles of the MW, which we show in the Appendix Fig.~\ref{fig:MW ColDensSubplt}. H{\,\sc i} is again influenced by individual satellite galaxies, and there is generally an increased scatter for ions tracing low-temperature gas in comparison to the high ions.

\subsubsection{Origins of the clumpy CGM at large radii}
\label{origins}

The presence of H{\,\sc i} sightlines mimicking Lyman limit-like column densities ($\sim 10^{17}$ cm$^{-2}$) in Fig.~\ref{fig:M31 ColDensSubplt} as well as Si{\,\sc iii} sightlines lying above $\gtrsim 10^{12}$ cm$^{-2}$, out to $R_{200}$ in our simulations, indicate that the cool-clumpy CGM extends to large distances up to virial radii. In fact, using VLT/UVES and  HST/STIS data, \citet{richter2003h2} as well as \citet{Richter_2009} have identified such a population of Lyman-limit like optical and UV absorption systems in the Milky Way halo at high radial velocities, most likely representing the observational counterpart of CGM clumps far away from the disk. It is then worthwhile to contemplate about the physical origins of this clumpy CGM gas. A comparison with corresponding H{\,\sc i} data from \citet{liangchen10.1093/mnras/stu1901} (henceforth, LC2014) reveals that the cool, clumpy CGM ($\log N($H{\,\sc i}$) > 10^{16}$ cm$^{-2}$) has a similar spatial extent (> 2--3 $R_\text{proj}/R_{200}$) as seen in our data. 

It is important to note that most of this clumpy CGM gas is not associated with the ISM of the satellite galaxies because those regions have far greater densities (a factor of $\sim$ 4-5 times higher) than that being discussed here. However, ram pressure stripping from the motions of many of the satellite galaxies within the $R_{200}$ of MW-M31 can deposit such intermediate-column density cool gas at these distances. We elaborate on ram pressure stripping and its effects on the CGM of LG in §~\ref{gas stripping}.

Gas accretion mechanisms onto the host galaxy, in itself could be a potential source for cool, slightly under-dense gas clumps manifesting as cold CGM at large distances. 

Galactic fountain flows have long been hypothesised as a possible means to efficiently circulate gas, metals and angular momentum between the ISM and the CGM \citep{fraternali2013ApJ...764L..21F, Fraternali2017}. Thermal instabilities arising from cold gas parcels from the ISM regions moving outward rapidly through the warm ambient CGM regions can result in the growth of intermediately dense cool gas. However, it is not immediately clear which of the above three processes could be the most dominant. While carrying out an elaborate tracer particle analysis or delving deeper into the ram pressure stripping processes could provide better clarity about the root cause of this distant cold CGM, it is beyond the scope of this paper.

\subsubsection{The bi-modal distribution of Si{\,\sc iii}}
\label{Bimodal}

Interestingly, Si{\,\sc iii} column density distributions show a strong bi-modality, with a higher sequence of sightlines clustered around $\sim 10^{14}$ cm$^{-2}$ and another lower sequence clustered around $\sim 10^{8}$ cm$^{-2}$. This bimodality is expected due to the bi-conical outflows, which we identified in Fig.~\ref{Fig:Skymaps}. This bimodal feature is indeed most prevalent for M31 in 09-18 and 17-11, where the bi-conical outflows were most visible. However, it is practically highly unlikely to detect the lower sequence of Si{\,\sc iii} column densities in near future; hence this bi-modal feature will not show up in the Si{\,\sc iii} observational datasets.

\subsection{Comparison with observations}
\label{Comparisons}

\begin{figure*}
\centering
\includegraphics[width=1.0\linewidth]{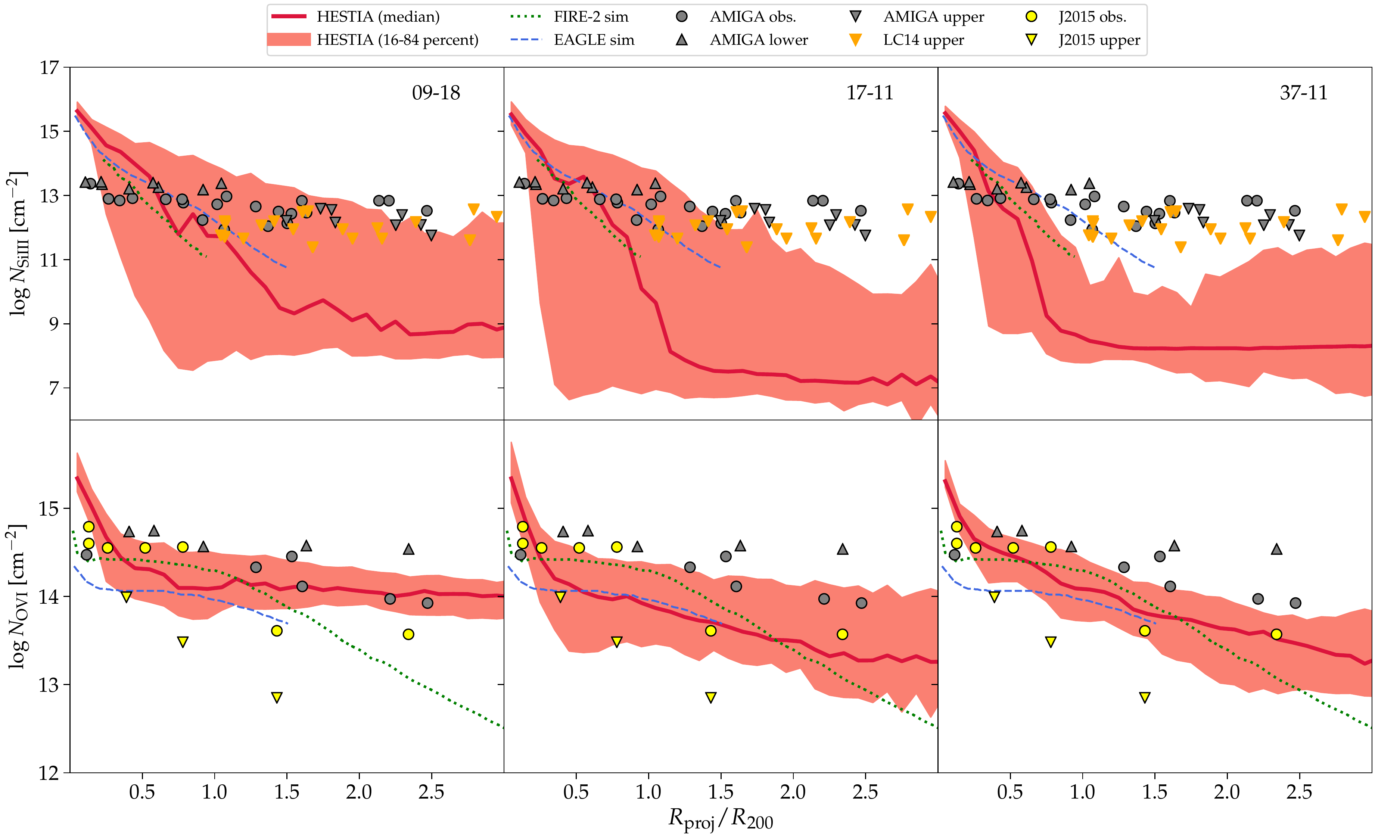}
\caption{The top (bottom) panel shows the Si{\,\sc iii} (O{\,\sc vi}) radial column density profiles for our three realizations for M31. The thick red curve denotes the median values while the red, shaded region denotes the 16-84th percentiles for our realizations. Circles refer to the detections while the downward and upward triangles, respectively, denote the upper and lower limits in the Project AMIGA survey \citep{lehner2020project}. The blue dashed line denotes the data from EAGLE simulations \citep{Oppenheimer2017TheMC} while the green dashed-dot line denotes the data from FIRE-2 simulations \citep{ji2020properties}. Downward orange triangles in the upper panel are Si {\,\sc iii} upper limits from the \citet{liangchen10.1093/mnras/stu1901}, while the yellow filled circles and downward triangles in the lower panel are O {\,\sc vi} measurements from \citet{johnson2015possible}. The Si{\,\sc iii} profile from \hestia is consistent with the LC14 upper limits, but there is an inconsistency between these and the Project AMIGA observations. Similarly, the J2015 and Project AMIGA observations of O{\,\sc vi} are inconsistent, and \hestia is only in reasonable agreement with J2015. In Fig.~\ref{Fig2005_ColDensMWcentered_SummaryPlot} we discuss that a likely explanation for the offset between \hestia and the AMIGA observations is contamination of gas from the MW to the AMIGA dataset.}
\label{SiIIIComparison}
\end{figure*}

While the previous subsections primarily dealt with the theoretical interpretations of our results from the power spectra and column density profiles, this subsection is dedicated to analysing how well these results match with data from observations and other simulations. We base our comparison on three different observational datasets:

\begin{itemize}
\item {\bf M31 observations from the Project AMIGA} \citep[Absorption Maps in the Gas of Andromeda; ][]{lehner2020project}. Project AMIGA is a UV HST program studying the CGM of M31 by using 43 quasar sightlines, piercing through its CGM at different impact parameters ($R_\text{proj}=25$ to $569$ kpc). Such a large number of sightlines for the Andromeda galaxy enables a constraining quantitative comparison to the corresponding mock data from our simulations. 
\item {\bf Absorption-line measurements of Si{\,\sc iii} from LC2014}. They present a study of low and intermediate ions in the CGMs of a sample of 195 galaxies in the low-redshift regime. However, 50\% of the LC2014 sample consists of dwarf galaxies. To enable a fair comparison, we select only galaxies in a comparable mass range to our M31 simulations. We specifically only include their galaxies with $10^{10.6}\; \text{M}_{\odot}\leq M_*\leq 10^{11.1} \;\text{M}_{\odot}$. In our context, the data pertaining to Si{\,\sc iii} (1206 $\Angstrom$) ion is relevant. Since this is an absorption-line study, they measure all ion abundances in terms of equivalent width (EW). In order to translate their EW measurements into column density values, we plot a corresponding curve of growth for different "$b$" parameters. From the curve of growth it is clear that for $\log N($Si{\,\sc iii}$)<12.0$ and $\log N($Si{\,\sc iii}$)>18.0$, translating EW into column densities is straightforward. However, in the $12.0<\log N($Si{\,\sc iii}$)<18.0$ regime, $b$-parameter degeneracy sets in and a single EW measurement can result in different values for column densities depending on the $b$-parameter adopted. For this reason we exclude the sightlines from LC2014 at distances $d/R_\text{200} < 1$ (where this degeneracy is present).
\item {\bf O {\,\sc vi} ion measurements from \citet*{johnson2015possible} (henceforth J2015)}. They present a study of distribution of heavy elements of sight-lines passing galaxies with different impact parameters. Like LC2014, the eCGM galaxy sample in J2015 also comprises of galaxies spanning a range of stellar masses (log \text{M}$_*$/\text{M}$_{\odot}$ = 8.4-11.5), \text{so we again apply a mass cut of} $10^{10.6}\; \text{M}_{\odot}\leq M_*\leq 10^{11.1} \;\text{M}_{\odot}$ and we also only include late-type galaxies.
\end{itemize}
In Fig.~\ref{SiIIIComparison} we compare the projected Si{\,\sc iii} and O{\,\sc vi} profiles for M31 from \hestia to these observational datasets, and we also show the EAGLE simulations \citep{Oppenheimer2017TheMC} and the FIRE-2 simulations \citep{ji2020properties}. We discuss the comparison to the other simulations in Sec.~\ref{SimComparisons}.

\subsubsection{Comparing \hestia to {\rm Si}{\,\sc iii} observations}

At low impact parameters, $R_\text{proj}\lesssim R_{200}$, the observed range of Si{\,\sc iii} column densities in AMIGA and our simulations are consistent\footnote{This is, of course, keeping in mind the uncertainties associated with the ion column densities in the innermost regions of the galaxies in our simulations i.e. regions where the ISM is dominant.}. For sightlines probing $R_\text{proj}\gtrsim 1.5 R_{200}$, our simulations under-predict the observed column densities. Some of the shown observational data points are upper limits, implying that the observations leave the possibility for individual sightlines with column densities as low as ours, but the simulations generally fall short by at least an order of magnitude at $R_\text{proj}\gtrsim 1.5 R_{200}$.

On the other hand, \hestia is perfectly consistent with the upper limits from LC2014. Indeed, there are tensions between the high Si{\,\sc iii} column densities reported by Project AMIGA (in M31) and the upper limits from LC2014. A possible reason for this could be contamination of gas from the MW halo or Local Group environment for the M31 observations. We will further assess this hypothesis in Sec.~\ref{MWcontamination}.

\subsubsection{Comparing \hestia to {\rm O}{\,\sc vi} observations}

When comparing the O {\,\sc vi} column density in AMIGA and \hestia we again see larger values in the former. At the same time, \hestia reveals larger column densities in comparison to the J2015 observations of galaxies from the low-redshift Universe.

The offset between the observed J2015 and AMIGA might again be caused by contamination of MW absorption in the latter dataset, or an alternative possibility for the offset is that one dataset is the mean of a sample of low-redshift galaxies and the other only takes into account a single galaxy's profile (M31).

\subsubsection{The normalization of the metallicity profile in \hestia}\label{MetallicityNorm}

In Appendix~\ref{RadialMetallicity} we show that the gas metallicity in the disc of the \hestia galaxies is up to a factor of 3 higher in comparison to observations. The na\"ive expectation is that the CGM gas metallicity is too high by a similar factor, and this would cause the \hestia column density profiles in Fig.~\ref{SiIIIComparison} to be overestimated by up to 0.5 dex. If we scale the \hestia M31 Si{\,\sc iii} and O{\,\sc vi} column density profiles down by 0.5 dex, the agreement with LC14 and J2015 improves, whereas the tension between the AMIGA observations and \hestia becomes stronger. This supports our conclusion that \hestia is well consistent with these observations of low-redshift galaxies.

\begin{figure*}
\centering
\includegraphics[width=1.0\linewidth]{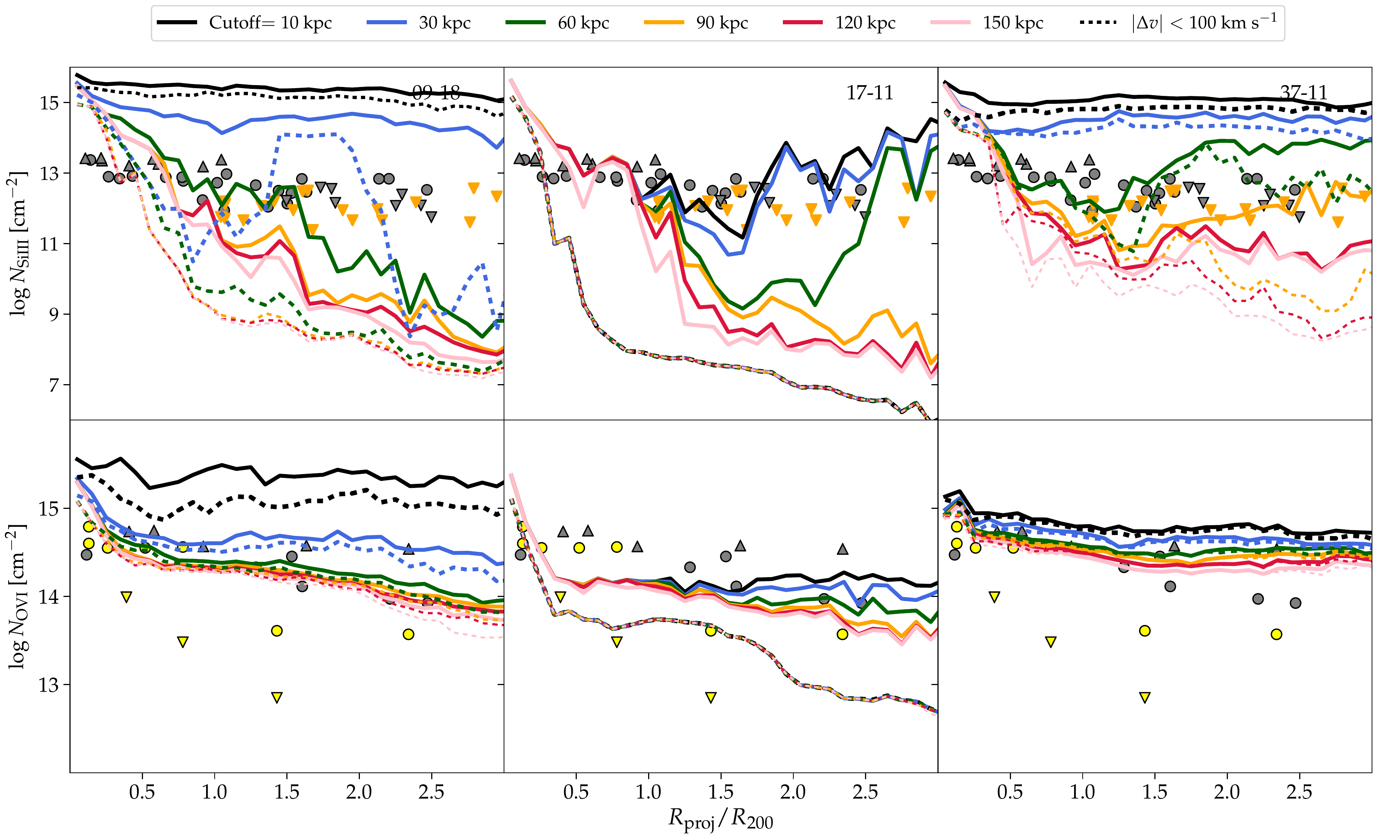}
\caption{We demonstrate how the gas in the MW's CGM may influence the observationally derived median column density profiles around M31. We have generated skymaps centred on the MW (instead of the LG, as done in previous figures), where we remove gas lying within a radial cutoff ranging from 10 to 150 kpc from MW (solid lines). For the dashed lines we additionally constrain gas to be within 100 km s$^{-1}$ of the M31. As in Fig.~\ref{SiIIIComparison}, the data points from Project AMIGA survey (filled grey markers), LC2014 (orange downward markers) and J2015 (filled yellow markers) have been overplotted. Even when only including gas within 100 km s$^{-1}$ of M31, the Si{\,\sc iii} profile of 09-18 and 37-11 is increased to $\simeq 10^{15}$ cm$^{-2}$ by clouds within 10--120 kpc from the MW centre. For 17-11, a velocity selection of gas very well removes gas within 150 kpc of the MW. For O{\,\sc vi}, the contamination from the MW's CGM is also significantly changing the profiles in 09-18 -- here gas residing within 150 kpc of the MW may boost the column density by 1.0 dex.}
\label{Fig2005_ColDensMWcentered_SummaryPlot}
\end{figure*}

\subsection{Comparison with other simulations}
\label{SimComparisons}

Fig.~\ref{SiIIIComparison} also shows the profiles for Si{\,\sc iii} and O{\,\sc vi} for EAGLE-based \citep{Oppenheimer2017TheMC} and FIRE-2 \citep{ji2020properties} based simulation datasets. For the comparison with EAGLE, we use the Si{\,\sc iii} and O{\,\sc vi} profiles from their $L_{\star}$ subsample, which has $\log M_{200} = 11.7-12.3 \; \text{M}_{\odot}$. It contains 10 haloes hosting star-forming galaxies. These are zoom simulations with non-equilibrium cooling. The corresponding average $R_{200}$ for this subsample is $\simeq 195$ kpc (see fig.~2 in \citealt{oppenheimer2016bimodality}). This dataset is at $z=0.2$, since the authors compare it with the COS-Halos data which covers the same redshift. For the FIRE-2 simulation comparison we compare to the m12i halo ($\log M_{200}\simeq 12$ M$_{\odot}$ at $z = 0$) using the FIRE-2 model with cosmic ray feedback (their simulation data is taken from fig.~17 in \citealt{lehner2020project}). Further details about the simulations and CGM modelling in FIRE2 simulations can be found in \citet{ji2020properties}.

The \hestia simulations show many similar trends to EAGLE and FIRE-2 and they, furthermore, \textit{all} under-predict the AMIGA column densities of Si{\,\sc iii} and O{\,\sc vi} at $R_\text{proj}\gtrsim 1.0 R_{200}$. On the other hand, all the simulations are broadly consistent with the observational datasets we have compiled based on LC2014 and J2015.

\subsection{Convergence test}
\label{Convergence}

In Appendix~\ref{SecConvergenceTest} we compare the high-resolution \hestia simulations analysed in Fig.~\ref{SiIIIComparison} with intermediate-resolution simulations having an eight times larger dark matter particle mass. This convergence test does not challenge our derived column density profile. 

Using the same simulation code and galaxy formation model as in our paper, \citet{van2019cosmological} showed that increasing the spatial resolution significantly boosts the H{\,\sc i} column density in the CGM. Idealised simulations furthermore reveal the possibility of gas to fragment to the cooling scales \citep{2018MNRAS.473.5407M,2019MNRAS.482.5401S}, which for dense gas is significantly below our resolution limit. Exploring the resolution requirement in the CGM of cosmological simulations is, however, still a field of ongoing research, so it is still a possibility that the idealized simulations over-estimate the needed spatial resolution.

We note that Si{\,\sc iii} and O{\,\sc vi} trace warmer gas in comparison to H{\, \sc i}, so these ions are expected to be less affected by resolution issues than H{\, \sc i}. Even though our convergence test does not reveal any signs of a lack of convergence, it is still a possibility that our column densities are affected by a too low spatial resolution.

\section{Discussion}
\label{Discussion}

\subsection{Biased column density profiles caused by the MW's CGM?}\label{MWcontamination}

We have found that observations of low-redshift galaxies disagree with the observed column densities of the M31 by the Project AMIGA. A possible explanation for this finding could be observational biases, for example, caused by gas clouds in the CGM of the MW contributing to the projected column density profile of M31. Such a bias does not play a role in our previous skymap analysis, because the skymaps are created by an observer in the geometric centre of the LG, and hence, the MW's CGM does not contribute to the sight-lines towards M31.

We now turn to addressing the role of such a bias in the three realizations of the \hestia simulations. We re-analyse our simulations with an observer located in the MW centre, and create skymaps of the different ions as before. In order to incorporate the larger distance from the MW to M31 (as opposed to the smaller distance from the LG centre to M31 in earlier analysis), we use longer sightlines (each 1400 kpc in length). To ensure grid-size uniformity with respect to the earlier analysis, we increase the number of gridpoints from 50,000 to 100,000. To determine the role of the MW's CGM, we create skymaps excluding gas within 10, 30, 60, 90, 120 and 150 kpc of the MW's centre. The corresponding projected radial column density profiles are seen as solid lines in Fig.~\ref{Fig2005_ColDensMWcentered_SummaryPlot}. The three different realizations show a significant amount of Si{\,\sc iii} and O{\,\sc vi} residing in the MW's CGM at a distance of 10--120 kpc from the MW's centre.

Observationally, a hint of the gas clouds' spatial origin can be obtained by looking at its line-of-sight velocity. In Fig.~\ref{Fig2005_ColDensMWcentered_SummaryPlot} we also construct profiles, where we exclude gas clouds with a line-of-sight difference ($|\Delta v|$) exceeding 100 km s$^{-1}$ from M31's velocity (see dashed lines in Fig.~\ref{Fig2005_ColDensMWcentered_SummaryPlot}). From our different realizations we see a different behaviour. For 09$-$18 and 37$-$11, the column density profiles of Si{\,\sc iii} and O{\,\sc vi} increase up to $10^{15}$ cm$^{-2}$ and by 1.0 dex, respectively (this is the difference between dashed lines indicating a cutoff of 10 kpc and 120 kpc in Fig.~\ref{Fig2005_ColDensMWcentered_SummaryPlot}), caused by gas residing between 10--120 kpc of the MW's CGM. For 17$-$11, the situation is less extreme, and the inferred column density profile of M31 is unaffected by the MW, when a velocity cut in the line-of-sight velocity is applied.

This analysis shows that the MW's CGM can substantially bias the inferred projected column densities of M31. For Si{\,\sc iii}, the potential bias is stronger in comparison to O{\,\sc vi}. For O{\,\sc vi} in 17$-$11, a velocity cut alone is successful in completely removing MW contributions. As seen from the lower middle panel in Fig.~\ref{Fig2005_ColDensMWcentered_SummaryPlot}, this still gives us a small discrepancy ($\sim$ 0.5 dex) with AMIGA observations. This means that our 17$-$11 analogues inherently do not produce enough O{\,\sc vi} to completely match the AMIGA observational trends. However, the opposite is true for the other two simulations where we clearly see our results matching fairly well with the AMIGA observations, when we include the contribution of gas from the MW halo. Overall, we infer that the biases estimated by our MW centred skymaps provide a likely explanation for the differences between the \hestia simulations and the AMIGA observations (seen in Fig.~\ref{SiIIIComparison}). At the same time, it also provides a likely explanation for the differences between the low-redshift galaxy samples (LC2014 and J2015) and the Project AMIGA\footnote{However, we do note here that both LC2014 and J2015 are a representative sample as opposed to the Project AMIGA observations which pertain to a single galaxy.}.

In reality, contamination of the gas from the Magellanic Stream (MS) to the M31 CGM observations is also a possibility. The MS passes just outside of the virial radius ($R_\text{vir} = 300$ kpc) of M31 (see fig.~1 in \citealt{lehner2020project}). For the purpose of ascertaining the level of MS contamination, \citet{lehner2020project} use Si{\,\sc iii} as their choice of ion (mainly because it is most sensitive to detect both weak as well as strong absorption). However, they do not remove entire sightlines merely on the suspicion of possible MS contamination. Instead, they analyze individual components and find that 28 out of 74 (38\%) Si{\,\sc iii} components are within the MS boundary region (and having Si{\,\sc iii} column density values larger than 10$^{13}$ cm$^{-2}$). These are not included in the sample from then on. For the remaining \emph{non-MS contaminated} components, they find a trend of higher Si{\,\sc iii} column density at regions away ($b_{\rm MS} > 15 ^\circ$) from the MS main axis ($b_{\rm MS} = 0^\circ$). This shows that the MS contamination is negligible for these components.

However, they do find a fraction (4/22) of dwarf galaxies out of their M31 dwarf galaxies sample falling in the MS contaminated region. This means that while they do take utmost care to avoid any MS contamination in their results, there could still be some residual contributions (especially in the cold gas observations of M31’s CGM) from the MW CGM. These could manifest in the form of slightly enhanced column densities in observations at regions beyond M31's virial radius.

\subsection{Gas stripping in the Local Group}
\label{gas stripping}

A characteristic that appears across all our realizations is the distorted nature of the CGMs of many satellite galaxies. High-velocity infall motions of dwarf galaxies through complex gravitational potential fields, typical in galaxy groups and clusters results in the dwarf galaxy CGM becoming structurally disturbed. In some extreme cases this can also result in trailing stripped gas tails \citep{smith2010ultraviolet, owers2012shocking, Salem_2015, mcpartland2016jellyfish, 2017ApJ...844...48P, Tonnesen2021}. While a few very clear examples of such galaxies have been described in detail in §\ref{Skymaps}, there are certainly many more. 

The role of stripped gas from the CGMs of satellite galaxies towards augmenting the pre-existing gas reserves of the host galaxy and thereby influencing the CGM of the host galaxy is rather well known from the observations of the MS, which emanates from the interaction of the Small and Large Magellanic Clouds on their approach towards the MW (e.g. \citealt{fox2014cos, Richter_2017}). However, a scarcity of deep observations means that very little is known about the part played by the diffuse gas from other satellite galaxies in our LG. Few studies pertaining to such observations reveal low neutral gas abundances around dwarf galaxies, though they might still harbour sizeable reserves of ionized gas \citep{westmeier2015neutral, emerick2016gas, fillingham2016under, simpson2018quenching}. 

By carefully analysing the gas flow kinematics across time-frames for these dwarf galaxies within \hestia, it will be possible, in future studies, to obtain not just their mock proper and bulk gas motions, but also various parameters regarding their stripped gas such as its spatial extent, cross-section and physical state. The Gaia DR2 proper motions of MW and LG satellites \citep{pawlowski2020milky}, along with corresponding comprehensive UV, optical and X-ray datasets from HST-COS, UVES, Keck and Chandra, can then provide us with clues regarding which \hestia realizations are most likely to produce these real observations. Furthermore, implementing similar sightline analysis, done in this paper for MW-M31, for multiple satellite systems over a range of their respective impact parameters, can yield extensive mock datasets that could then prove useful in the wake of future surveys that will be sensitive to even lower column density gas.

\subsection{Physical modelling of the CGM}\label{TensionsM31}

In recent years, our understanding of the CGM has dramatically improved, and it is encouraging that our simulations are broadly consistent with observations. This is despite of our relatively simple physics model.

Theoretical work has for example suggested that parsec-scale resolution, which is so far unattainable in cosmological simulations like \hestia, may be necessary to resolve the cold gas in galaxies \citep[][-- we note, however, that these results are so far only suggestive and the need for parsec-scale resolution has so far not been demonstrated, yet this could be a potential reason for the offset]{2018MNRAS.473.5407M,2019MNRAS.482.5401S,2019ApJ...882..156H, van2019cosmological}. Results from \citet{van2019cosmological} proved that $\sim$ 1 kpc resolution in the CGM boosts small-scale cold gas structure as well as covering fractions of Lyman limit systems; this might also hold true for slightly less dense but slightly more ionized cool gas. \citet{mccourt10.1093/mnras/stx2687} proposed a cascaded shattering process via which a large cloud experiencing thermal instability can cool a couple of orders of magnitude (from $T \sim 10^{6}$ K to $\sim 10^{4}$ K), mainly as a result of continued fragmentation within the larger cloud. They compute the characteristic length scale, associated with shattering, to be $\sim 1-100$ pc. Multiple observations also show that cool gas is indeed present in form of small clouds out to $\sim R_\text{vir}$ in galaxy haloes \citep{lau2016quasars,hennawi2015Sci...348..779H,stocke2013characterizing,prochaska2008quasars}. 
Using Cloudy ionization models, \citet{Richter_2009} have determined the characteristic sizes of the partly neutral CGM clumps in the MW halo based on their HST/STIS absorption survey, leading to typical scale lengths in the range 0.03 to 130 pc (see tables 4 \& 5 in \citealt{Richter_2009}). From their absorber statistics, these authors estimated that the halos of MW-type galaxies contain millions to billions of such small-scale gas clumps and argue that these structures may represent transient features in a highly dynamical CGM. Thus, it is clear that the length scales involved in these processes are still at least an order of magnitude below what is currently achievable in the highest resolution zoom-in simulations. It is also worth mentioning that \citet{2021arXiv210805355F} have recently introduced a novel framework modelling multiphase winds, which may be relevant for future cosmological simulations of the CGM.

\citet{lehner2020project} discusses feedback processes, which may also affect how gas and metals are transported to large radii. The role of cosmic ray feedback in influencing the CGM has recently gained interest from multiple research groups \citep{Salem_2014, salem2016role, buck2020effects, ji2020properties, hopkins2020}, and it has been shown to significantly alter gas flows in the CGM of simulations. CR-driven winds from the LMC \citep{bustard2020ApJ...893...29B} as well as those from the resolved ISM \citep{Simpson_2016, girichidis2018, Farber_2018} have been shown to change both the outer and inner CGM properties, respectively. Similarly, magnetic fields have been shown to influence the physical properties of the CGMs of simulated galaxies, thereby modifying the metal-mixing in the CGM \citep{vandeVoort2021}.

Despite of \hestia agreeing relatively well with the observations, we note that there are still some important challenges for future galaxy formation models in terms of understanding physical processes in the CGM.

\section{Conclusions}
\label{Conclusions}

We have analysed the gas, spanning a range of temperatures and densities, around the MW-M31 analogues at $z = 0$ in a set of three \hestia simulations. These LG simulations use the quasi-Lagrangian, moving mesh {\sc arepo} code, along with the comprehensive Auriga galaxy formation model. We have set our frame of reference to the LG geometrical centre and generated ion maps for a set of five ions, H{\,\sc i}, Si{\,\sc iii}, O{\,\sc vi}, O{\,\sc vii} and O{\,\sc viii}. Some important conclusions have emerged from our study: 

\begin{itemize}

\item We have created mock skymaps of the gas distribution in the LG. All dense gas blobs with $N_\text{HI}>10^{20}$ cm$^{-2}$ are associated with a galaxy; either a satellite galaxy or MW/M31 themselves. The skymaps of H{\,\sc i} and Si{\,\sc iii} reveal strong imprints of satellite galaxies, whereas the tracers of warmer gas (O{\,\sc vi}, O{\,\sc vii} and O{\,\sc viii}) are mainly dominated by the haloes of MW and M31. The projected column density profiles of the latter ions are, indeed, well-described by monotonic decreasing functions of the impact parameter. In comparison, the projected H{\,\sc i}- and Si{\,\sc iii}-profiles have a much higher scatter caused by blobs associated with the satellite galaxies.
\item A power spectrum analysis of the skymaps shows that H{\,\sc i}, Si{\,\sc iii}, O{\,\sc vi}, O{\,\sc vii} and O{\,\sc viii} have a gradually higher coherence angle on the sky -- ions tracing the coldest gas are most clumpy. This confirms the impression we get by visually inspecting the skymaps, and it is also consistent with the behaviour of the column density profiles.
\item The visual inspection of the simulated skymaps reveal multiple satellite galaxies with disturbed gas morphologies, especially in H{\,\sc i} and Si{\,\sc iii}. These are LG analogues of jellyfish galaxies. Future simulation analyses and observations can give a unique insight to the physical processes in the ISM and CGM of these galaxies. 
\item For the \hestia M31 analogues we compare the Si{\,\sc iii} and O{\,\sc vi} column density profiles to observations of M31 and low-redshift galaxies. 
The spectroscopic observations of M31 and low-redshift galaxies reveal remarkably different column density profiles. Using our simulations, we find that the gas residing in the Milky Way may contaminate the sight-lines towards M31, such that the M31 column densities are boosted. For Si{\,\sc iii} and O{\,\sc vi} we see this contamination boosting the column density profiles up to as much as $10^{15}$ cm$^{-2}$ and by 1.0 dex, respectively, even when only including gas within a 100 km s$^{-1}$ of the M31 velocity. Contamination of gas from the MW, hence, provides one of the likely explanations for the offset between observations of M31 and low-redshift galaxies.
\item The M31 analogues from \hestia have  Si{\,\sc iii} and O{\,\sc vi} column density profiles broadly consistent with low-redshift galaxy constraints. If we include a contamination from MW gas, then in 2 out of 3 M31 realizations we can also reproduce the large column densities observed in the direction of M31 in Project AMIGA.

\end{itemize}

\section*{Data availability}
The scripts and plots for this article will be shared on reasonable request to the corresponding author. The \textsc{arepo} code is publicly available \citep{weinberger2020arepo}.

\section*{Acknowledgements}
We thank Moritz Itzerott, Martin Wendt, Gabor Worseck for useful comments and discussions. We also thank the anonymous referee for some very constructive comments which greatly improved the quality of this paper. MS acknowledges support by the European Research Council under ERC-CoG grant CRAGSMAN-646955. MHH acknowledges support from William and Caroline Herschel Postdoctoral Fellowship Fund. SEN is a member of the Carrera del Investigador Cient\'{\i}fico of CONICET. He acknowledges support by the Agencia Nacional de Promoci\'on Cient\'{\i}fica y Tecnol\'ogica (ANPCyT, PICT-201-0667). RG acknowledges financial support from the Spanish Ministry of Science and Innovation (MICINN) through the Spanish State Research Agency, under the Severo Ochoa Program 2020-2023 (CEX2019-000920-S). MV acknowledges support through NASA ATP grants 16-ATP16-0167, 19-ATP19-0019, 19-ATP19-0020, 19-ATP19-0167, and NSF grants AST-1814053, AST-1814259,  AST-1909831 and AST-2007355. ET acknowledges support by ETAg grant PRG1006 and by EU through the ERDF CoE grant TK133. The authors sincerely acknowledge the Gauss Centre for Supercomputing e.V. (\url{https://www.gauss-centre.eu/}) for providing computing time on the GCS Supercomputer SuperMUC at the Leibniz Supercomputing Centre (\url{http://www.lrz.de/}) for running the \hestia simulations. We thank the contributors and developers to the software packages \textsc{yt} \citep{2011ApJS..192....9T} and \textsc{astropy} \citep{2018AJ....156..123A}, which we have used for the analysis in this paper.

%%%%%%%%%%%%%%%%%%%%%%%%%%%%%%%%%%%%%%%%%%%%%%%%%%


\bibliographystyle{mnras}
\bibliography{BibMNRAS} % if your bibtex file is called example.bib

%%%%%%%%%%%%%%%%%%%%%%%%%%%%%%%%%%%%%%%%%%%%%%%%%%

%%%%%%%%%%%%%%%%% APPENDICES %%%%%%%%%%%%%%%%%%%%%
\appendix

\section{Radial gas metallicity profiles}\label{RadialMetallicity}

\begin{figure*}
    \centering
    \includegraphics[width = \linewidth]{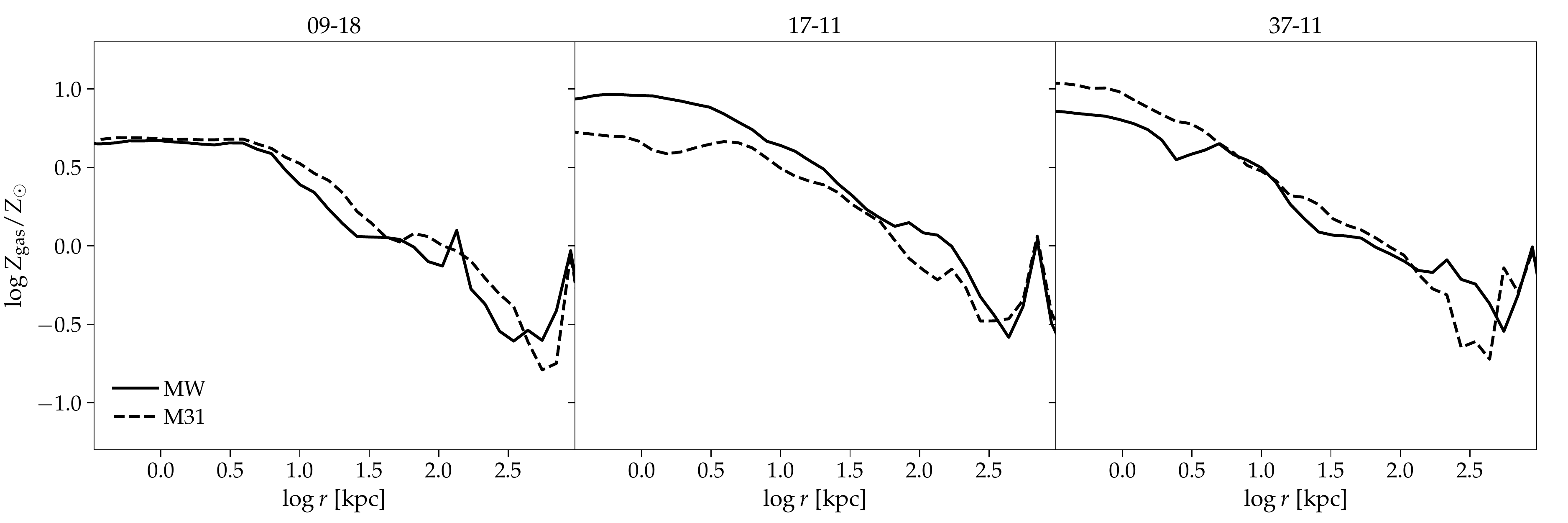}
    \caption{Radial gas metallicity profiles for the \hestia galaxies. The profiles show two distinct regimes-- metal-rich in the inner disc regions ($r < 10$ kpc) and metal-poor in the CGM regions ($r > 30$ kpc). The rise in metallicities at $r > 500$ kpc occurs due to the presence of the pairing galaxy at these distances.}
    \label{fig:Metallicity}
\end{figure*}

We obtain the radial gas metallicity profiles in spherical shells equally spaced in the logarithmic radius ($\log r$) for the \hestia galaxies in Fig.~\ref{fig:Metallicity}. Overall, the gas metallicities for MW and M31 look similar. The \hestia galaxies are metal-rich in the inner disc regions (3-10 times the solar metallicity inside 10 kpc), after which the metallicity drops sharply out to the CGM regions (as low as 0.2 times solar metallicity at 500 kpc). Beyond this point, the metallicities rise again due to the presence of the pairing galaxy at those distances. As observed by \citealt{Conroy_2019}, we also see our galaxies exhibiting a turn-over from being metal-rich (at $r < 10$ kpc) to metal-poor (at $r > 30$ kpc).

For the MW in 17-11 and M31 in 37-11 the central gas metallicities reach values as high as $10 Z_\odot$. These values are clearly a factor of 2-3 higher than for M31 observations \citep{sanders2012ApJ...758..133S}, and these also exceed our expectations for MW-like galaxies (see fig. 10 in \citealt{2014MNRAS.438.1985T} for a compilation of observations of MW-mass galaxies). We, therefore, conclude that \hestia produces a disc metallicity, which is up to a factor of 3 higher than expected from observations. There are no strong observational constraints on the MW and M31 CGM metallicity, but when comparing to observations we keep in mind the possibility that our simulations might have a CGM metallicity, which is up to a factor of 3 too high in comparison to \emph{real galaxies}.

\section{A listing of the most relevant parameters for the most massive galaxies in each realization}
\label{Listing}

In Table~\ref{tab:0918Catalog},~\ref{tab:1711Catalog}~and~\ref{tab:3711Catalog} we show properties of the satellite galaxies in each of the simulations. The galaxy numbers appear in Fig.~\ref{Fig:Skymaps} of the main paper, and we see that all the dense H{\,\sc i} regions are associated with one of the galaxies listed in the tables.

\begin{table*}
\centering
\caption{A list of properties for the most massive galaxies in the 09$-$18 realization. Galaxy no. 0 corresponds to the M31, while galaxy no. 9 corresponds to the MW. Remaining galaxies can be correlated with their respective galaxy nos. in Fig.~\ref{Fig:Skymaps}. Dist. (kpc) refers to the distance of the corresponding galaxy from the LG centre, in kpc.}
\label{tab:0918Catalog}
\begin{tabular}{ccccc} 
		\hline
		Galaxy no. & log $ M_*$ (M$_{\odot}$) & log $M_\text{dm}$ (M$_{\odot}$) & log $M_\text{gas}$ (M$_{\odot}$) & Dist. (kpc)\\
		\hline
		\hline
		0 & 11.113 & 12.275 & 11.195 & 433.19\\
		\hline
		1 & 9.184 & 10.338 & 9.627 & 494.37\\
		\hline
		2 &  8.755 & 10.164 & 9.385 & 585.21\\
		\hline
		3 & 8.445 & 10.033 & 9.258 & 622.04\\
		\hline
		4 & 8.807 & 9.614 & 8.902 & 474.55\\
		\hline
		5 & 8.973 & 9.038 & 8.666 & 420.03\\
		\hline
		6 & 8.665 & 9.540 & 8.636 & 478.16\\
		\hline
		7 & 7.854 & 9.729 & 8.453 & 328.25\\
		\hline
		8 & 8.189 & 9.098 & 7.599 & 335.97\\
		\hline
		9 & 10.911 & 12.156 & 11.078 & 433.19\\
		\hline
		10 & 10.390 & 11.111 & 10.207 & 554.78\\
		\hline
		11 & 9.220 & 10.345 & 9.707 & 658.60\\
		\hline
		12 & 8.896 & 10.254 & 9.596 & 767.30\\
		\hline
		13 & 9.026 & 9.952 & 9.503 & 415.00\\
		\hline
		14 & 8.640 & 10.133 & 9.419 & 547.98\\
		\hline
		15 & 8.718 & 9.748 & 9.207 & 525.90\\
		\hline
		16 & 8.012 & 9.907 & 8.967 & 549.84\\
		\hline
		17 & 7.983 & 9.771 & 9.025 & 152.61\\
		\hline
		18 & 8.638 & 9.469 & 8.839 & 420.42\\
		\hline
		19 & 6.909 & 9.152 & 8.151 & 572.19\\
		\hline
		20 & 7.010 & 9.237 & 7.800 & 684.53\\
		\hline
		21 & 7.769 & 9.052 & 7.444 & 508.79\\
		\hline
		22 & 5.203 & 9.241 & 5.246 & 421.25\\
		\hline
		23 & 8.181 & 9.944 & 9.138 & 387.82\\
		\hline
		24 & 7.261 & 9.910 & 8.785 & 683.50\\
		\hline
		25 & 7.389 & 9.822 & 8.931 & 641.68\\
		\hline
		26 & 7.497 & 9.556 & 8.419 & 658.64\\
		\hline
	\end{tabular}
\end{table*}

\begin{table*}
\centering
\caption{Same as \ref{tab:0918Catalog}, but for the 17$-$11. Galaxy no. 0 corresponds to the M31, while galaxy no. 1 corresponds to the MW.}
\label{tab:1711Catalog}
\begin{tabular}{ccccc} 
		\hline
		Galaxy no. & log $ M_*$ (M$_{\odot}$) & log $M_\text{dm}$ (M$_{\odot}$) & log $M_\text{gas}$ (M$_{\odot}$) & Dist. (kpc)\\
		\hline
		\hline
		0 & 11.079 & 12.310 & 11.212 & 338.01\\
		\hline
		1 & 11.062 & 12.184 & 10.919 & 338.00\\
		\hline
		2 & 9.648 & 10.330 & 9.422 & 455.76\\
		\hline
		3 & 8.985 & 10.429 & 9.697 & 259.56\\
		\hline
		4 & 8.686 & 10.282 & 9.745 & 244.43\\
		\hline
		5 & 9.461 & 10.074 & 9.414 & 306.94\\
		\hline
		6 & 8.745 & 10.304 & 9.593 & 640.84\\
		\hline
		7 & 8.977 & 10.023 & 9.619 & 114.45\\
		\hline
		8 & 9.334 & 9.702 & 8.899 & 389.40\\
		\hline
		9 & 8.469 & 9.654 & 9.248 & 233.49\\
		\hline
		10 & 8.195 & 9.843 & 9.044 & 409.12\\
		\hline
		11 & 8.029 & 9.246 & 8.753 & 240.47\\
		\hline
		12 & 7.028 & 9.620 & 8.369 & 263.22\\
		\hline
		13 & 6.871 & 9.401 & 8.495 & 422.59\\
		\hline
		14 & 7.285 & 9.435 & 8.344 & 336.52\\
		\hline
		15 & 7.446 & 9.407 & 8.141 & 400.67\\
		\hline
		16 & 6.860 & 9.416 & 7.818 & 499.36\\
		\hline
		17 & 7.486 & 5.304 & 8.611 & 324.31\\
		\hline
		18 & 6.725 & 5.605 & 8.539 & 359.37\\
		\hline
		19 & 6.036 & - & 7.750 & 383.76\\
		\hline
		20 & 7.894 & 9.880 & 8.906 & 676.74\\
		\hline
		21 & 7.563 & 9.801 & 8.635 & 696.55\\
		\hline
		22 & 7.364 & 9.631 & 8.607 & 491.8\\
		\hline
		23 & 7.617 & 9.571 & 8.240 & 788.02\\
		\hline
		24 & 6.538 & 9.460 & 8.057 & 466.96\\
		\hline
		25 & 6.772 & 9.498 & 7.998 & 678.25\\
		\hline
		26 & 5.953 & 9.460 & 7.092 & 496.14\\
		\hline
	\end{tabular}
\end{table*}

\begin{table*}
\centering
\caption{Same as \ref{tab:0918Catalog}, but for the 37$-$11. Galaxy no. 0 corresponds to the M31, while galaxy no. 11 corresponds to the MW.}
\label{tab:3711Catalog}
\begin{tabular}{ccccc} 
		\hline
		Galaxy no. & log $ M_*$ (M$_{\odot}$) & log $M_\text{dm}$ (M$_{\odot}$) & log $M_\text{gas}$ (M$_{\odot}$) & Dist. (kpc)\\
		\hline
		\hline
		0 & 10.719 & 11.955 & 10.871 & 425.30\\
		\hline
		1 & 8.919 & 10.599 & 9.809 & 484.08\\
		\hline
		2 & 9.299 & 10.349 & 9.799 & 584.03\\
		\hline
		3 & 8.956 & 10.209 & 9.678 & 471.87\\
		\hline
		4 & 8.246 & 10.297 & 9.315 & 517.37\\
		\hline
		5 & 7.086 & 9.974 & 8.205 & 637.63\\
		\hline
		6 & 8.380 & 9.493 & 8.765 & 475.21\\
		\hline
		7 & 7.640 & 9.530 & 8.772 & 635.12\\
		\hline
		8 & 7.046 & 9.439 & 8.489 & 518.32\\
		\hline
		9 & 6.647 & 9.031 & 7.611 & 510.55\\
		\hline
		10 & 6.593 & 7.969 & 6.532 & 519.30\\
		\hline
		11 & 10.774 & 11.954 & 10.761 & 425.29\\
		\hline
		12 & 9.517 & 10.706 & 9.876 & 473.75\\
		\hline
		13 & 7.153 & 9.533 & 8.198 & 353.56\\
		\hline
		14 & 6.882 & 9.357 & 7.818 & 317.15\\
		\hline
		15 & 6.934 & 9.267 & 7.870 & 567.53\\
		\hline
		16 & 7.669 & 8.939 & 8.092 & 325.65\\
		\hline
		17 & 6.536 & 9.223 & 6.542 & 516.28\\
		\hline
		18 & 10.040 & 11.377 & 10.343 & 645.13\\
		\hline
		19 & 8.356 & 9.781 & 9.202 & 611.27\\
		\hline
		20 & 8.326 & 9.339 & 8.767 & 683.16\\
		\hline
		21 & 7.515 & 9.440 & 8.630 & 705.46\\
		\hline
		22 & 7.133 & 9.487 & 8.331 & 504.48\\
		\hline
		23 & 9.675 & 10.895 & 9.841 & 698.75\\
		\hline
		24 & 8.470 & 10.121 & 9.270 & 728.57\\
		\hline
		25 & 7.693 & 10.049 & 8.944 & 638.65\\
		\hline
		26 & 7.411 & 9.746 & 8.482 & 230.93\\
		\hline
		27 & 7.151 & 9.581 & 8.348 & 576.87\\
		\hline
		28 & 5.682 & 9.580 & 7.388 & 744.04\\
		\hline
		29 & 6.073 & 9.515 & 6.705 & 426.66\\
		\hline
		30 & 6.534 & 9.465 & 7.609 & 731.14\\
		\hline
		31 & 6.536 & 9.204 & 6.780 & 574.73\\
		\hline
	\end{tabular}
\end{table*}

\section{Column density profiles for the MW}

\begin{figure*}
    \centering
    \includegraphics[width = \linewidth]{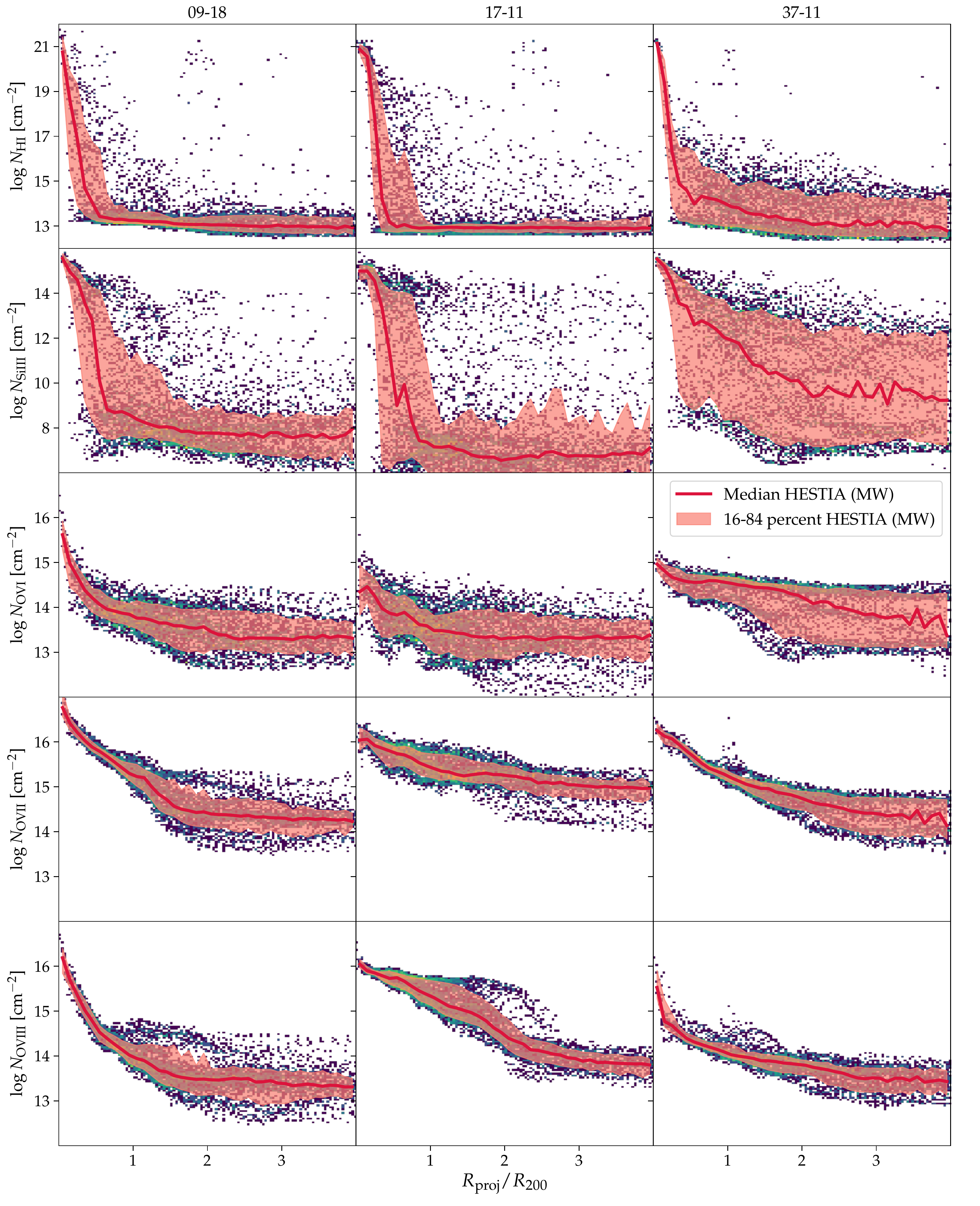}
    \caption{Same as Fig.~\ref{fig:M31 ColDensSubplt}, but for MW. A distinct blob of H{\,\sc i} column density absorbers, which can be seen at a distance of $\sim$ 2.0 $R_\text{200}$ in the H{\,\sc i} profile for 09$-$18, can be correlated with the satellite galaxy numbered 17 in the corresponding skymap (H{\,\sc i} skymap for 09$-$18 in Fig.~\ref{Fig:Skymaps}).}
    \label{fig:MW ColDensSubplt}
\end{figure*}

In Fig.~\ref{fig:MW ColDensSubplt} we show the radial column density profile of the simulated MW for the different ions. This is complementary to the M31 column density profiles in Fig.~\ref{fig:M31 ColDensSubplt}.

\section{Convergence test}\label{SecConvergenceTest}

\begin{figure*}
\centering
\includegraphics[width=1.0\linewidth]{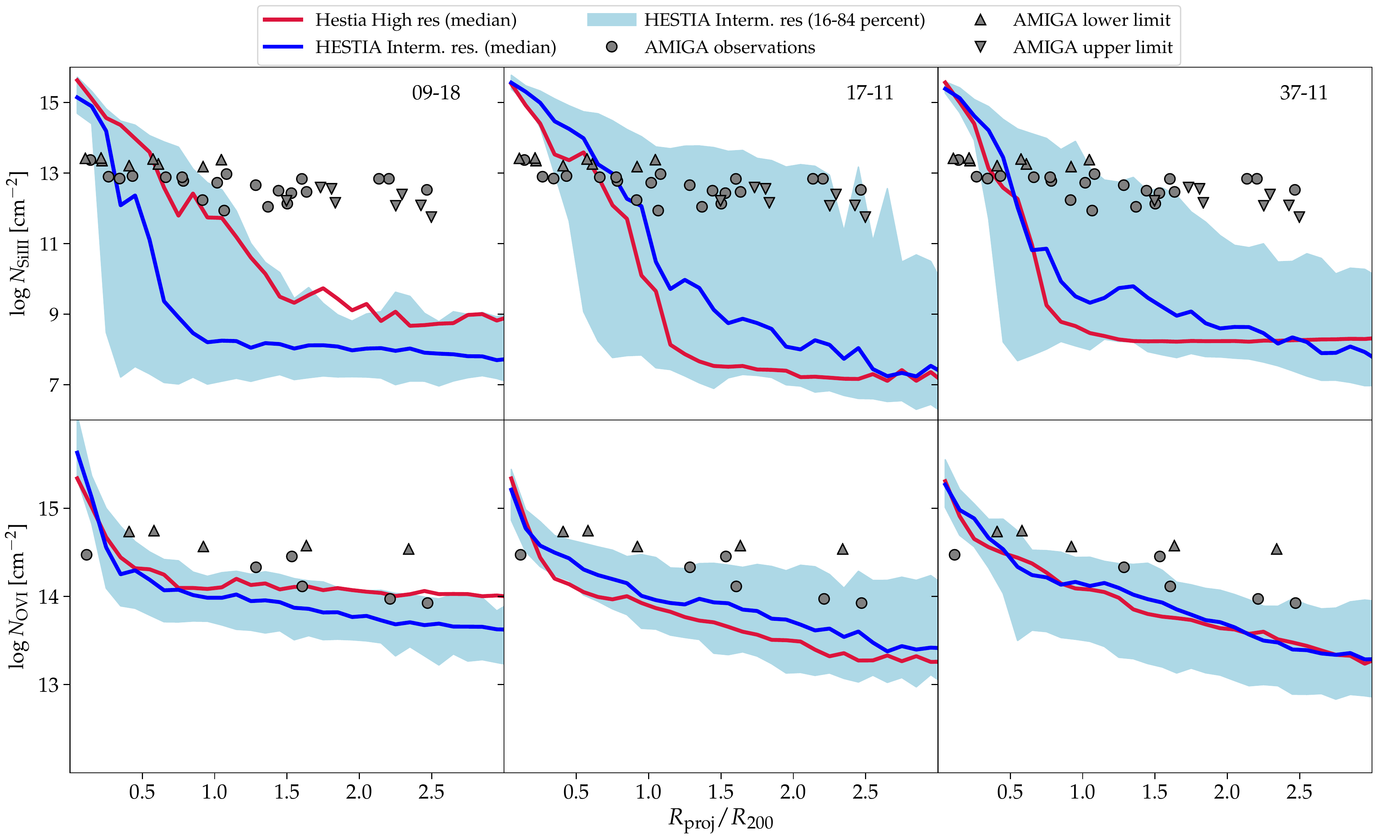}
\caption{We perform a convergence test of Fig.~\ref{SiIIIComparison}. The thick red line shows the median of the high-resolution \hestia simulations, which was also shown in Fig.~\ref{SiIIIComparison}. The blue line and contour show the median and 16-84 percentiles, respectively, of intermediate resolution simulations with an eight times lower mass resolution (dark matter particles have an eight times higher mass) in comparison to the high resolution simulations. Examination of the median profiles does not indicate a lack of convergence, so our column density profiles are well converged.}
\label{Percentiles_M31ConvergenceTest}
\end{figure*}

We perform a convergence test, where we compare the high-resolution \hestia simulations, which we presented in the main paper, to intermediate-resolution simulations with an eight times lower mass of the dark matter particles. In Fig.~\ref{Percentiles_M31ConvergenceTest}, we test whether the column density profiles of Si{\,\sc iii} and O{\,\sc vi} are converged. In simulation 09-18, the column densities at $\gtrsim R_{200}$  are higher in the intermediate resolution simulation in comparison to the high-resolution simulations. For 17-11 and 37-11, we have the opposite trend -- we see the highest column densities in the high-resolution simulations. The median profiles of O{\,\sc vi} are only slightly affected by resolution with the difference between intermediate and high resolution simulations being less than a factor of two. We conclude that, on the whole, the column density profiles are well converged.

% Don't change these lines
\bsp	% typesetting comment
\label{lastpage}
\end{document}